# Quantitative electrochemical control over optical gain in quantum-dot solids

Jaco J. Geuchies[†], Baldur Brynjarsson[†], Gianluca Grimaldi[†*], Solrun Gudjonsdottir[†], Ward van der Stam[†$], Wiel H. Evers[†], Arjan J. Houtepen[†]

[†] Optoelectronic Materials Section, Faculty of Applied Sciences, Delft University of Technology, Van der Maasweg 9, 2629 HAZ Delft, The Netherlands.

[*] Current address: Center for Nanophotonics, AMOLF, Science Park 104, 1098 XG Amsterdam, The Netherlands.

[$] Current address: Inorganic Chemistry and Catalysis, Debye Institute for Nanomaterials Science, Utrecht University, Universiteitsweg 99, 3584 CG Utrecht, The Netherlands.

ABSTRACT

Solution processed quantum dot (QD) lasers are one of the holy-grails of nanoscience. They are not yet commercialized because the lasing threshold is too high: one needs > 1 exciton per QD, which is hard to achieve due to fast non-radiative Auger recombination. The threshold can however be reduced by electronic doping of the QDs, which decreases the absorption near the band-edge, such that the stimulated emission (SE) can easily outcompete absorption. Here, we show that by electrochemically doping films of CdSe/CdS/ZnS QDs we achieve quantitative control over the gain threshold. We obtain stable and reversible doping more than two electrons per QD. We quantify the gain threshold and the charge carrier dynamics using ultrafast spectroelectrochemistry and achieve quantitative agreement between experiments and theory, including a vanishingly low gain threshold for doubly doped QDs. Over a range of wavelengths with appreciable gain coefficients, the gain thresholds reach record-low values of $\sim 10^{-5}$ excitons per QD. These results demonstrate an unprecedented level of control over the gain threshold in doped QD solids, paving the way for the creation of cheap, solution-processable low-threshold QD-lasers.

INTRODUCTION

Colloidal semiconductor quantum dots (QDs) are attractive materials for solution-processable and color-tunable lasers[1–4]. Additionally, having discrete electronic states with finite degeneracy, QDs are ideal systems to achieve low threshold optical gain, promising a reduction of the threshold current in lasers. However, the larger-than-unity degeneracy of the conduction and valence-band levels implies that multiexcitons are required to achieve population inversion in QDs, i.e. for stimulated emission (SE) to outcompete absorption. Since Auger recombination is efficient in QDs[5], multiexcitons have short lifetimes, typically < 100ps, implying that it is difficult to achieve/maintain a population sufficient for gain. Additionally, ultrafast charge-carrier trapping can compete with the build-up of optical gain when the trapping rates are similar to the time for the QDs to achieve population inversion[6].

To overcome these limitations, prolonged Auger lifetimes up to a nanosecond have been achieved in QD heterostructures.[7,8] Moreover, in CdSe QDs, the band-edge hole degeneracy can be decreased in the presence of strain[9], reducing the number of excitons needed to achieve population inversion to ~1. Sub-single-exciton optical gain has been realized in Type-II heterostructures[10] and in HgTe QDs[11], which show a large shift of the SE to wavelengths where there is little absorption, at the cost of lower gain coefficients.

A potentially more controllable method to suppress absorption employs QD charging[12–14]. In 2004, pioneering work on electrochemical charging by the group of Guyot-Sionnest showed a reduction in the threshold for amplified stimulated emission (ASE)[15]. Recently, Wu et al. demonstrated nearly thresholdless optical gain using photochemical doping as a strategy to charge the QDs[7], which, when coupled to a distributed feedback grating, shows sub-single exciton lasing.[16]. While these recent results show the promise of QD charging for lasing, there is limited and only temporary control over the charge density and as a result, thresholdless gain has not been achieved. Electrochemical doping has the advantage that the electrochemical potential can be fixed and held constant over a QD film. When the QDs are sufficiently passivated with stable shells, this results in a stable and homogeneous doping density of the QD film.[17,18]

In the current work we seek to get quantitative understanding and control over the gain properties of QD films when doped with electrons. We combined spectroelectrochemistry with ultrafast transient absorption (TA) spectroscopy, characterizing material gain as a function of electrochemical doping density and optical excitation density. We measured the number of photogenerated excitons per QD to produce optical gain as a function of the average number of electrochemically injected electrons into the 1S(e) conduction band state. In a broad wavelength range, we achieve vanishingly low optical gain thresholds ($< 10^{-3} - 10^{-5}$ excitons per QDs). Modelling the effect of state filling, stimulated emission and carrier relaxation on optical gain, we get good agreement between the predicted and experimentally determined gain threshold and gain lifetime as a function of $<n_{1S(e)}>$. This demonstrates that we have quantitative control over the optical gain in these QD solids.

RESULTS AND DISCUSSION

We synthesized wurtzite core-shell-shell CdSe/CdS/ZnS QDs for the experiments presented here. Details regarding the synthesis[19–21] and characterization are presented in the Methods section. The steady-state absorption and photoluminescence (PL) spectra, and a representative transmission electron microscopy image are shown in Figure 1(a). The epitaxial shells increase the absorption cross-section of the QDs at the excitation wavelength, boost the electrochemical stability of the QDs[17] and lead to a PL quantum yield of 81% in solution (see Methods).

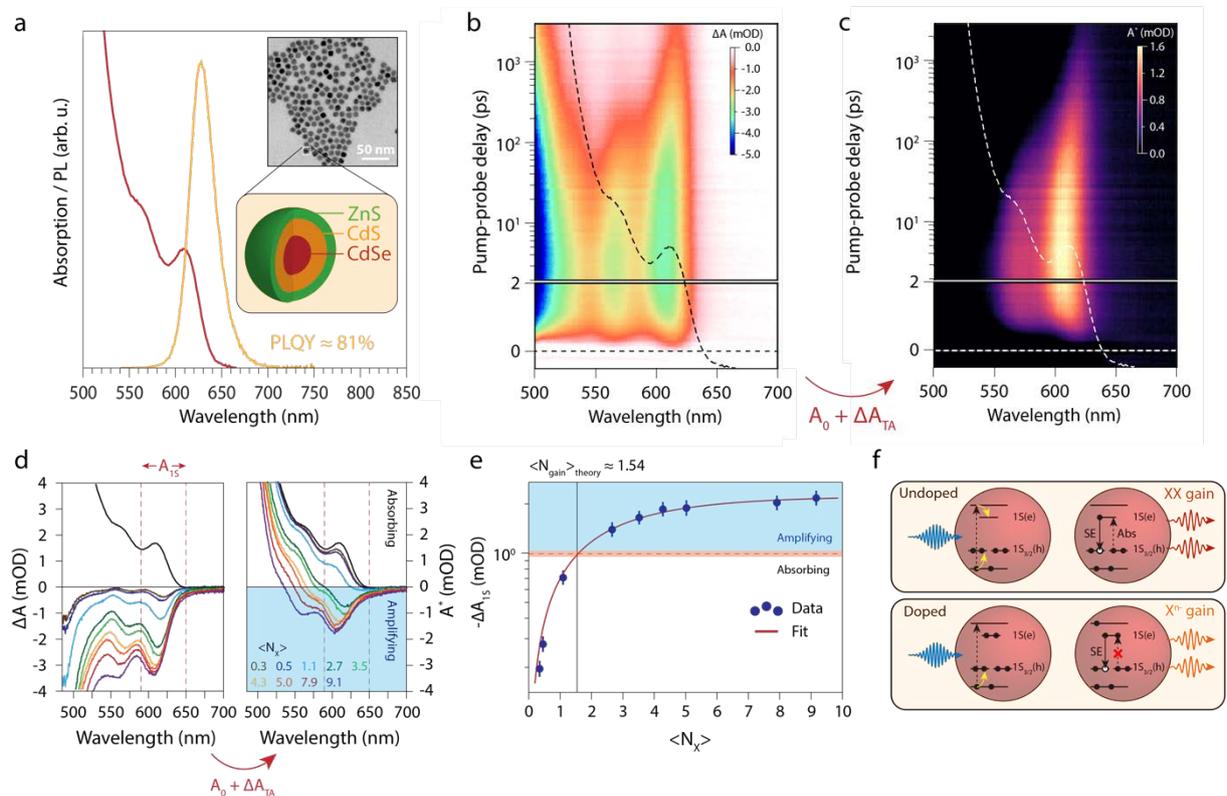

**Figure 1: Benchmarking the neutral CdSe/CdS/ZnS QDs in solution.** (a) Steady-state absorption and PL spectra of the CdSe/CdS/ZnS QDs used throughout this work. The insets show a TEM micrograph of the QDs, which have a diameter of 10.9±1.0 nm, and a schematic of the QDs. (b) 2D TA image for a pump wavelength of 400 nm and excitation density of $\langle N_X \rangle$ = 7.9. The black dotted line shows the steady-state absorption spectrum. (c) Time-dependent excited state absorption spectrum (or gain-map), obtained from (b), where all the positive absorption is colored black, and the negative absorption (gain) is colored following the scale

bar. **(d)** Spectral slices at a pump-probe delay time of 5 ps for increasing pump-fluence. The left image displays the differential absorption spectra, whereas the right image shows the excited-state absorption spectra. **(e)** Quantification of the optical gain threshold in solution, showing -ΔA (@ 5ps pump-probe delay time) versus <$N_X$>. The red line indicates a heuristic fit (increasing exponential function) to the data, used to determine the gain threshold. **(f)** Schematic of the mechanism for low threshold optical gain in undoped and doped QDs. Upon above-bandgap-excitation, both the electron and hole will cool down to the band-edges. In the n-doped QDs, the 1S absorption transition is already blocked. Hence, stimulated emission does not have to compensate for the absorption process, which leads to zero-threshold optical gain.

To benchmark these QDs we first measure the optical gain threshold in solution. We determine the absorption cross section ($\sigma = 3.6 \pm 0.2 \cdot 10^{-14}$ cm$^2$ at the 400 nm pump wavelength) of the QDs, from the fluence dependence of Auger recombination [6,22] (see SI). Using the measured photon fluence J, we determine the average number of photogenerated excitons per QD, <$N_X$> = $J\sigma$. Figure 1(b) shows a colormap of a TA measurement on the QDs. By adding the steady-state absorption, $A_0$, to the transient absorption, we obtain a gain map, as shown in Figure 1(c), showing in color the region of the spectrum characterized by negative absorption in the excited state, i.e. optical gain. To quantify the gain spectra and gain threshold, we take spectral slices at 5 ps, after thermalization of hot carriers, which is shown in Figure 1(d). The gain of the lowest the energy transition ((1S(e) – 1S$_{3/2}$(h), the 1S transition) starts redshifted compared to the steady-state absorption spectrum, but shows a distinct blueshift with increasing excitation fluence (to a maximum of 20 meV). This is in agreement with models by Bisschop et al.[23], who showed that the biexciton binding energy becomes repulsive when thick CdS shells are grown around the QDs.

The electronic transitions leading to optical gain in intrinsic and doped QDs are schematically depicted in Figure 1(f). After photoexcitation above the band-gap of the CdSe core, hot carriers rapidly cool to the conduction and valence band-edges of the QD. Once thermalized, a second photon with an energy equal to the band-gap energy can either lead to absorption or stimulated emission. Depending on the excitation density, the QDs either remain absorptive, become transparent, or show optical gain.

To quantitatively evaluate the optical gain, we spectrally average the TA spectrum over the band-edge transition at 5 ps time delay [dashed vertical lines in Fig 1(d)]. Comparing the averaged bleach, presented in Figure 1(e), with the average absorption over the same wavelength range (horizontal dashed line), we determined the 1S gain-threshold $<N_{gain,1S}>$ to be 1.55±0.07 excitons per QD. This is in quantitative agreement with the theoretical value of $<N_{gain,1S}>$ of 1.54 for a two-fold 1S(e) and four-fold $1S_{3/2}$(h) degeneracy, determined considering a Poissonian distribution of excitons over the QD [24] (see SI, section 1.1). Furthermore, we observe that the absorption of the 1S transition is completely inverted at 5 ps for the highest pump fluence ($<N_X>$ = 9.1). At high pump fluences ($<N_X> \geq 3.5$), also the second transition (1S(e) – $2S_{3/2}$(h), the 2S transition) shows optical gain. These results show that the neutral QDs behave nearly ideally and their gain properties are understood quantitatively.

# ELECTROCHEMICAL DOPING OF QD FILMS

To quantify the relationship between optical gain and electronic doping, we need precise control over the doping density inside the QD film. Using spectroelectrochemical (SEC) measurements, we controllably inject carriers into the QD films, monitoring changes in the photoluminescence and absorption of the film to determine the doping density. We prepared QD-films by spincoating a QD dispersion in toluene on a conductive ITO-on-glass substrate, followed by crosslinking the QDs with 1,7-diaminoheptane, to ensure the films have a good electron mobility (see Methods).

Figure 2(a) shows SEC differential absorption ($\Delta A_{SEC}$) measurements. We sweep the potential between the open circuit potential (-0.30V vs. the Ag pseudoreference electrode (PRE), i.e. -0.77V vs. Fc/Fc$^+$, see SI) and -1.50V (i.e. -1.97V vs. Fc/Fc$^+$), while measuring the change in absorption of the QD film. Upon electron injection into the conduction band of the QD film, we observe a decrease of several absorption transitions as a result of state filling of the 1S(e) conduction band level. Figure 2(a) shows three electrochemical cycles, highlighting the excellent reversibility of $\Delta A_{SEC}$. $\Delta A_{SEC}$-spectra at selected potentials are shown in Figure 2(b). The magnitude of the band-edge bleach is equal to the ground state absorption ($A_0$) at roughly -1.4V, indicating that the 1S(e) level is completely filled. Figure 2(c) shows the corresponding SEC photoluminescence spectra as a function of applied potential. The PL intensity drops severely upon electron injection into the conduction band, an expected consequence of increased Auger decay in the n-doped QDs[17]. Figure 2(d) shows PL spectra at different potentials.

Figure 2(e) shows the absorption spectra of the charged QD films, i.e. the sum of the $\Delta A_{SEC}$-spectra and the ground state absorption spectrum. The 1S transition becomes transparent at -1.4V.

To quantify the charge density, we fitted a Gaussian to the 1S absorption bleach feature at every potential, as well as to the 1S feature in the ground state absorption spectrum. The average number of electrochemically injected electrons in the 1S(e) level is given by $\langle n_{1S(e)}\rangle = 2\cdot\Delta A_{1S(e)}/A_{0,1S(e)}$, where we use the Gaussian amplitudes of the fitted 1S absorption and absorption bleach[13]. The resulting values of $\langle n_{1S(e)}\rangle$ at each potential are shown, together with the normalized PL intensity, in Figure 2(f). Charging and discharging of the QD film is fully reversible, as the number of electrochemically injected electrons into the 1S(e) level oscillates between zero and two. Furthermore, we observe that the absorption bleach increases at the same potential as the PL starts to quench, a good indication of trap-free and electrochemically stable QDs[17,25].

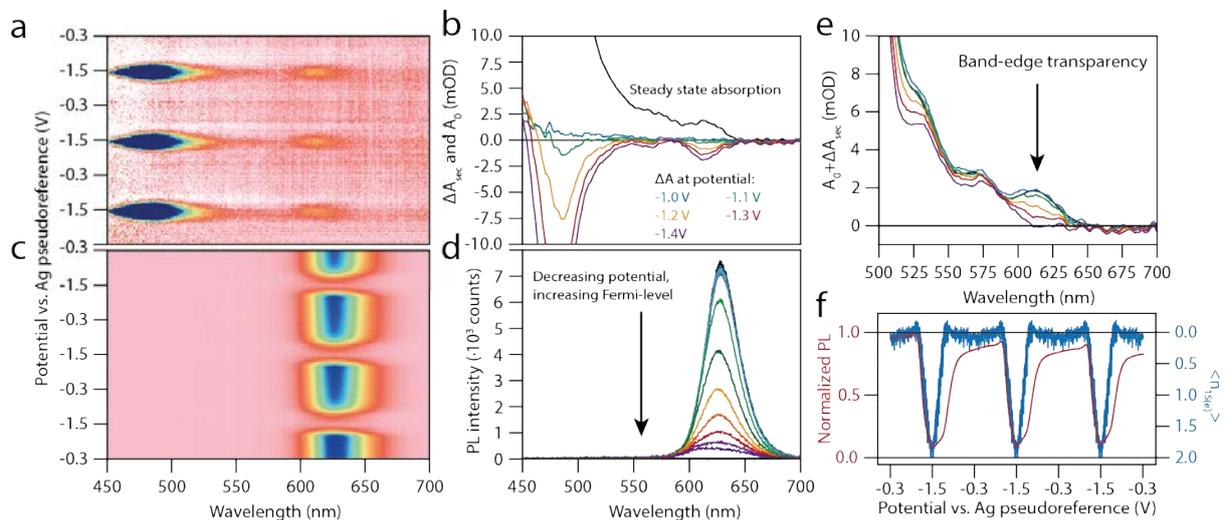

**Figure 2: Spectroelectrochemistry on a film of CdSe/CdS/ZnS QDs.** The potential during all SEC measurements was swept three times between the open circuit potential and -1.5V to check for sample stability. **(a)** SEC absorption measurements. Injection of electrons into the conduction band of the QDs is observed by bleaching of the band-edge (around 615 nm) and CdS shell (< 550 nm) transitions. **(b)** $\Delta A_{SEC}$ spectra at different potentials. Charge injection

starts around -1.1V. Note that the band-edge bleach amplitude at the most negative potentials equals the amplitude of the absorption spectrum. **(c)** SEC PL measurements. As electrons are injected into the conduction band of the QDs, the PL quenches due to Auger recombination. **(d)** PL spectra at different potentials. The PL amplitude decreases due to enhanced Auger recombination. **(e)** Total absorption of the QD film, i.e. $\Delta A_{SEC}+A_0$. The band-edge transition becomes transparent at -1.4V. **(f)** Normalized PL and amplitude of the band-edge bleach as a function of applied potential. The drop in PL coincides with the injection of charges into the conduction band of the QDs, indicating a relatively trap-free QD film. The number of electrochemically injected 1S(e) electrons oscillates between zero and two.

ULTRAFAST SPECTROELECTROCHEMISTRY

To characterize the gain response of n-doped QD films, we performed fs transient absorption (fsTA) measurements while electrochemically controlling the doping density, which we refer to as ultrafast spectroelectrochemistry. The differential absorption signal, $\Delta A_{TA}$, can be added to the steady-state absorption spectrum of the sample to obtain the excited state absorption. The gain threshold is defined as the first excitation fluence resulting in a negative excited state absorption.

Figure 3 presents excited state absorption spectra for 400 nm excitation as a function of pump-probe delay time for various electrochemical doping densities ranging from $<n_{1S(e)}>$ = 0 to 2 and excitation fluences ranging from $<N_X>$ = 0 to 6.6. As optical excitation of a thin film of semiconductor material results in changes in both the absorption and reflection of the film, all TA spectra are corrected for changes in reflectivity of the sample after photoexcitation[26]. The procedure for correcting the as-measured 'transient-extinction' signal of the QD film is outlined in the SI.

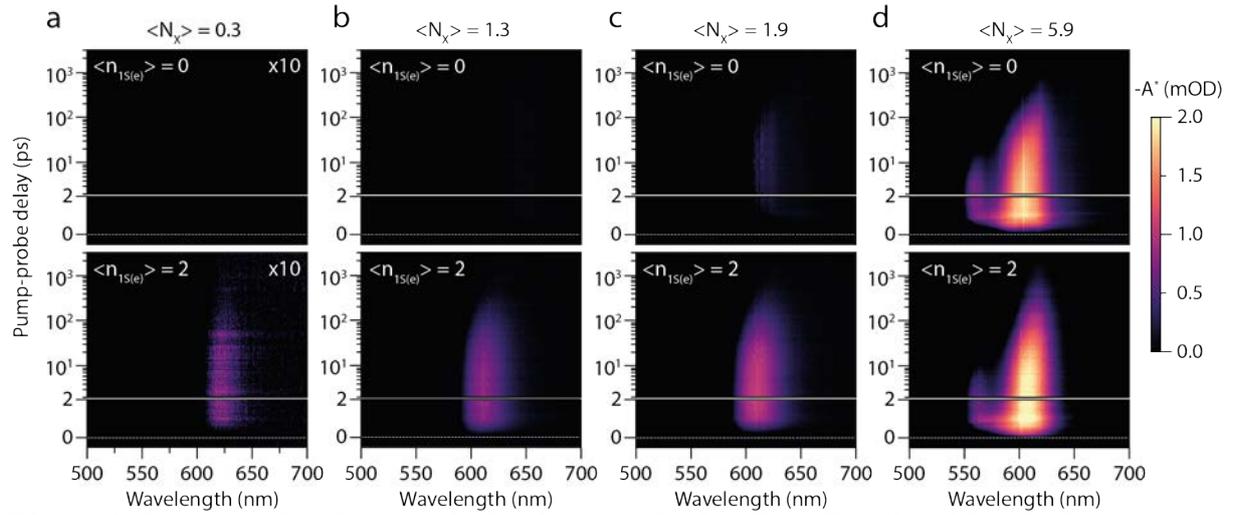

**Figure 3: Reduced threshold optical gain upon doping the QD solid with two electrons per QD.** Excited state absorption maps (excitation at 400 nm) as a function wavelength and pump-probe delay time. The upper panels show absorbance of the undoped QD solid ($<n_{1S(e)}> = 0$), whereas the bottom panels show the doped QD solid ($<n_{1S(e)}> \sim 2$). **(a)** The low fluence data ($<N_X> = 0.3$) were multiplied by 10 for clarity. The doped QD solid already shows optical gain around 620 nm at the lowest excitation fluence. **(b)** For $<N_X> = 1.3$, slightly below the theoretical threshold of $<N_{gain}> = 1.54$ for the undoped QDs, no signature of optical gain is observed in the undoped QDs, whereas the gain amplitude is increased for the doped solid. **(c)** For $<N_X> = 1.9$ a small amount of optical gain for the undoped QDs is observed. **(d)** For high photoexcitation density, resulting in $<N_X> = 5.9$, both the undoped and doped QD solid show full inversion of the 1S transition, and a shorter-lived gain signal originating from the 2S transition.

The top row of Figure 3 shows gain maps for the undoped film, i.e. $<n_{1S(e)}> = 0$ (-0.3V vs Ag. PRE), whereas the bottom row shows gain maps are for $<n_{1S(e)}> = 2$ (at -1.5V vs. Ag PRE). From left to right the excitation density increases. Figure 3(a) shows the excited state absorption map for $<N_X> = 0.3$. For clarity, the signal amplitude is multiplied by 10. For the undoped QD film we do not observe any optical gain over the measured spectral window. Upon increasing the excitation fluence to $<N_X> = 1.9$, we start to observe gain from the 1S transition in Figure 3(c). At the highest fluence presented here, $<N_X> = 5.9$, we also observe optical gain from the 2S transition. In stark contrast, at a doping density of $<n_{1S(e)}> = 2$, we observe light amplification of the 1S transition even for the lowest excitation fluence, as shown in the bottom panel of Figure 3(a). Upon increasing the fluence, the optical gain amplitude increases, and we again observe optical gain from the 2S transition at $<N_X> = 5.9$. The data qualitatively show that electrochemical doping can drastically reduce the optical gain threshold. In the remainder of the manuscript we will quantify the relationship between optical gain, density of electrochemically injected carriers and density of excitons.

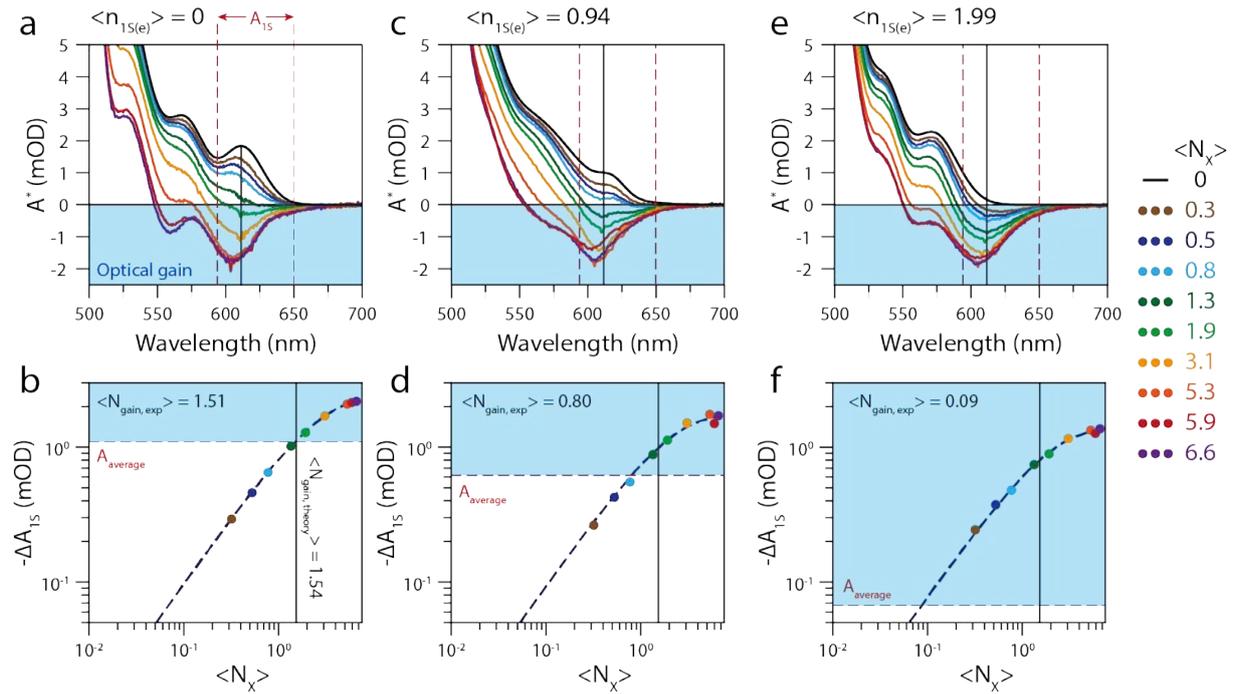

**Figure 4: Gain threshold determination for different doping densities.** Spectra were recorded at an excitation wavelength of 400 nm and a pump-probe delay time of 5 ps. **(a)** Excited state absorption spectra at open circuit potential, where $<n_{1S(e)}> = 0$, and for varying excitation densities. At $<N_X> = 1.9$, we start to see negative absorption. For the two highest excitation densities, we also observed optical gain originating from the 2S transition. The dashed red lines show the region in between which the absorption (and bleach) amplitude is averaged. **(b)** Bleach amplitude ($-\Delta A_{1S}$) versus the excitation density $\langle N_X \rangle$ for the 1S transition in the uncharged QD solid. The dashed red line shows the average absorption of the 1S transition above which the QD film shows optical gain. The fitted gain threshold of $\langle N_{gain,1S} \rangle$ = 1.51 agrees well with the theoretically expected gain threshold of 1.54 excitons per QD. **(c)** Excited state absorption spectra at a doping density of $\langle n_{1S(e)} \rangle$ = 0.94. **(d)** Determination of the gain threshold for $\langle n_{1S(e)} \rangle$ = 0.94. The steady-state absorption at the band-edge transition is reduced, resulting in $\langle N_{gain,1S} \rangle$ = 0.80. **(e)** Excited state absorption spectra at a doping density of $\langle n_{1S(e)} \rangle$ = 1.99. **(f)** Determination of the gain threshold for $\langle n_{1S(e)} \rangle$ = 1.99. The band-edge absorption transition is transparent due to the electrochemically injected electrons, reducing $\langle N_{gain,1S} \rangle$ to 0.09 excitons per QD.

For every doping density, we determine the average number of excitons required to reach transparency of the averaged 1S transition, $\langle N_{gain,1S} \rangle$. The data for the undoped film, and the film doped with $\langle n_{1S(e)} \rangle$ = 0.94 and $\langle n_{1S(e)} \rangle$ = 1.99 are presented in Figure 4. The upper panels of the figure show the excited-state absorption (i.e. $A_0 + \Delta A_{SEC} + \Delta A_{TA}$) at a pump-probe delay time of 5 ps. The bottom panels show the gain-threshold determination (similar to Figure 1(e)).

The gain properties of the undoped film (Figures 4(a), (b)) are nearly identical to those of the QDs in solution, shown in Figure 1. We determine a spectrally averaged gain threshold of $\langle N_{gain,1S} \rangle$ = 1.51. The the QD film with a doping density $\langle n_{1S(e)} \rangle \sim 1$, shows a reduced threshold of $\langle N_{gain,1S} \rangle$ = 0.80 (Figures 4(c) and 4(d)). At the highest doping density, $\langle n_{1S(e)} \rangle \sim 2$ (Figures 4(e) and 4(f) the optical gain threshold for the integrated 1S transition is reduced to $\langle N_{gain,1S} \rangle$ =

0.09 ± 0.09 excitons/QD. For all doping densities, increasing the fluence leads to saturation of the optical gain, reaching similar maximum values for different doping densities.

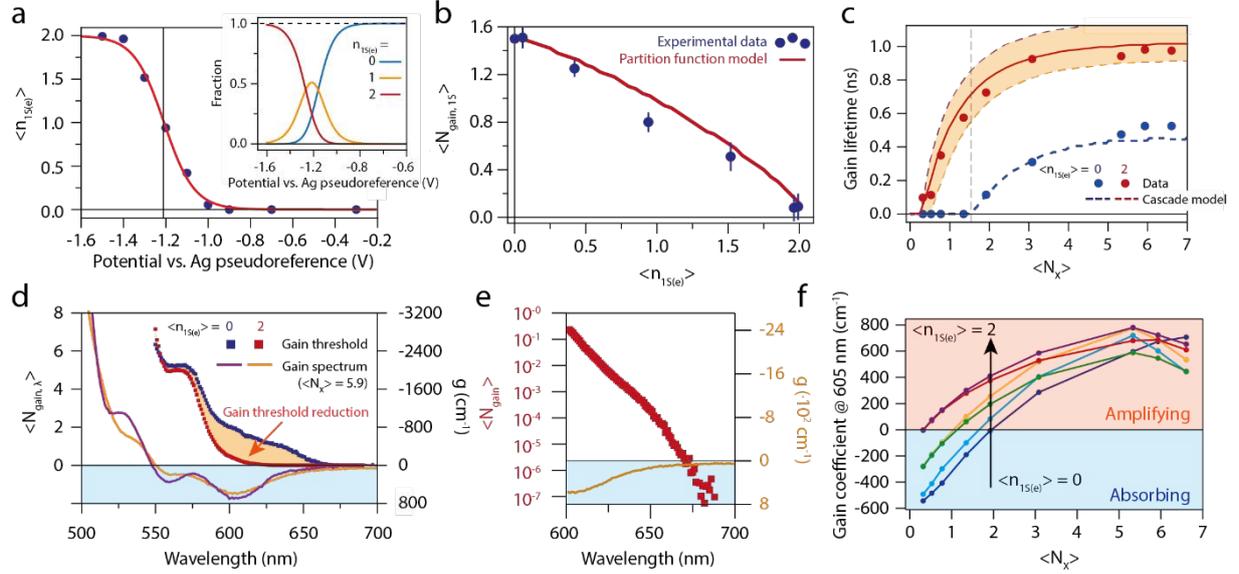

**Figure 5: Quantification and modelling of measured optical gain and determination of device-relevant parameters.** (a) $<n_{1S(e)}>$ versus the applied electrochemical potential. Blue datapoints represent the experimentally determined charge density, the red solid line is a Fermi-Dirac fit. The inset shows the distribution of neutral, singly-charged and doubly-charged QDs *vs*. applied potential as extracted from the fit. (b) Gain threshold $<N_{gain,1S}>$ versus the average number of electrochemically injected electrons per QD $<n_{1S(e)}>$. The blue datapoints represent the experimentally determined threshold, the solid red line is a model based on Poisson statistics for photon excitation, Fermi-Dirac statistics for electron filling, and transition counting to estimate the absorption cross section of the band-edge transition. (c) Gain lifetimes as a function of $<N_X>$ for the neutral and doubly charged QD film. The blue and red dots are experimental data points, and the lines represent a model based on coupled rate-equations (with $<n_{1S(e)}>$ = 1.95, 1.9 and 1.85 going from top to bottom). (d) Single wavelength gain threshold $<N_{gain,\lambda}>$ *vs*. wavelength for the neutral and $<n_{1S(e)}>$ = 2 doped QD film, shown as the blue and red squares respectively. For comparison, we also show the excited state absorption spectrum for the neutral and charged film for $<N_X>$ = 5.9 (blue and red solid lines respectively), to demonstrate that the optical gain threshold vanishes at wavelengths where there is an appreciable gain coefficient.

**(e)** a zoom-in of **(d)** the spectral region from 600-700 nm, plotted on a logarithmic scale. We observe $<N_{gain,\lambda}> = 10^{-3}–10^{-5}$ in regions with significant negative absorption. **(f)** Gain coefficient at 605 nm for various doping densities. The gain coefficients saturate around 800 cm$^{-1}$.

Finally, we combine all experiments shown above to develop a model that quantitatively describes optical gain in doped QDs. We model the excited state absorption in a QD film using the transition model counting model discussed in the SI, section 1.1-1.5, and in ref [6]. As inputs we need to know the distribution of electrons and excitons over the QDs at each potential. First, fitting a Fermi-Dirac distribution for a two-fold degenerate level to $<n_{1S(e)}>$ as a function of the applied potential, shown in Figure 5(a), we obtain the fraction of neutral, singly charged and doubly charged QDs at each potential (see SI, Section 1.3). Combined with a Poisson distribution of $<N_X>$ at each fluence this allows us to predict the gain threshold $<N_{gain,1S}>$ as a function of $<n_{1S(e)}>$. The modelled excited-state absorption is:

$$A^*_{ensemble}(\langle N_X \rangle, \langle n_{1S(e)} \rangle) = \sum_{n_{echem}=0}^{2} \sum_{N=0}^{N_{max}} f(\langle n_{1S(e)} \rangle) \cdot P(\langle N_X \rangle, N) \cdot A^*(N, \langle n_{1S(e)} \rangle) \quad (1)$$

with $P(\langle N_X \rangle, N)$ a Poisson distribution for the exciton density and $f(\langle n_{1S(e)} \rangle)$ a Fermi-Dirac function describing electrochemical state-filling of the 1S(e) level (see SI, section S1). Equation 1 is numerically solved to determine the value $\langle N_X \rangle$ where $A^*_{ensemble} = 0$. The prediction from this model is shown as the red solid line in Figure 5(b).

The data points in Figure 5(b) show the measured gain threshold $<N_{gain,1S}>$ as a function of $<n_{1S(e)}>$, showing a decrease in threshold with increasing $<n_{1S(e)}>$. Note that the red solid line is not a fit to the data, but a predication based on independent experimental observables (i.e. the absorption cross section from fluence dependent Auger recombination data (Figure S12 in the SI), and the Fermi-Dirac distribution from SEC-absorption measurements, Figure 5(a)). The match between the experimental gain thresholds and the model prediction is excellent, demonstrating that we have quantitative control over the gain threshold, both experimentally and theoretically.

In Figure 5(c) we present the lifetime of the optical gain signal for the neutral QD film ($<n_{1S(e)}>$ = 0, blue datapoints), and the QD film with on average two electrons per QD (red datapoints). We define the optical gain lifetime as the amount of time after photoexcitation that the average excited state 1S absorption remains negative. The extracted gain lifetimes increase with both increasing $<N_X>$ and with increasing $<n_{1S(e)}>$. For undoped QD films the highest gain lifetime is ~0.5 ns, while for doped QD it reaches ~1 ns. To model the gain dynamics, we set up a system of coupled differential equations that take Auger decay of (charged) multiexcitons into account (see SI, section S1.5). From the fluence-dependent fsTA data, we extract a biexciton lifetime of 310 ps (see SI). We assume that Auger rates scale with the number of electrons and holes as outlined by the group of Klimov[5,27,28], which allows us to model the cascade of Auger processes controlling excitonic decay. As shown in Figure 5(c), we get good quantitative agreement with the experimental data.

These results demonstrate the possibility to use our analytical model to accurately describe the relationship between the carrier population and optical gain in QD solids. Furthermore, complementing the transition counting model with a description of excitonic decay allows a precise prediction of the gain lifetime. Having been validated on this dataset, the model provides useful insight on the lasing characteristics of different QD materials of known band-edge degeneracies and Auger lifetimes, and can be used to direct the development of novel devices.

So far, we have focused on the spectrally averaged gain threshold $<N_{gain,1S}>$, as this is most insightful to study the effects of state filling on absorption and stimulated emission, without complications from spectral shifts that result from doping or optical excitation. However, for practical applications, a more relevant value is the gain threshold at a single wavelength, where light amplification is to take place, $<N_{gain,\lambda}>$. In Figure 5(d) we plot $<N_{gain,\lambda}>$ as a function of wavelength for the undoped (blue squares) and the $<n_{1S(e)}> = 2$ charged (red squares) QD film. The decrease of the optical gain threshold is clearly visible for the 1S transition, highlighted with the yellow area in the graph. For comparison, the gain coefficient spectra (at a fluence of $<N_X> = 5.9$) are also plotted in the same figure. These are obtained from the excited state absorption spectra and the film thickness d = 116 ± 13 nm (see Methods), as $g = \frac{-A^* \ln(10)}{d}$.

Figure 5(e) shows a zoom of Figure 5(d), but on a logarithmic scale. We observe that the threshold practically vanished for the doped film over a significant wavelength range. It becomes clear that, defined at a single wavelength, the gain threshold is somewhat arbitrary. In addition to a low threshold it is important that there is a significant gain coefficient at the amplified wavelength, at least large enough to compensate for losses that occur for the optical mode that is amplified. A typical loss coefficient in InGaAs/GaAs/AlGaAs laser diode arrays >50 cm$^{-1}$.[29] Taking this gain coefficient as the threshold for amplification we observe a record-low single-wavelength thresholds of 2.5·10$^{-5}$ excitons per QD, corresponding to a 400 nm pump fluence of 0.4 nJ/pulse/cm².

Figure 5(f) shows the gain coefficient at 605 nm as a function of $<N_X>$ and for various doping densities. At a fixed value of $<N_X>$ the gain coefficient is always significantly higher for doped QD films than for the neutral film. The maximum gain coefficient for the doped QD film is ~800 cm$^{-1}$, which is similar to the intrinsic gain coefficient of colloidal QDs in solution and III-V epitaxial semiconductors (10$^3$ cm$^{-1}$).[23,30] This demonstrates the great promise of electrochemically doped QD films for use as low-threshold gain media with strong light amplification. The next step, currently underway in our lab, is to employ electrochemically doped QD films in devices such as DFB gratings[16,31–34,31–34], micro-disk lasers[35], and ring resonators[36].

CONCLUSIONS

We have demonstrated precise experimental and theoretical control over the optical gain threshold in QD solids, via controlled and reversible electrochemical doping. After electrochemically injecting on average two electrons per QD into the 1S(e) electron level, we show that the spectrally integrated 1S gain threshold is as low as 0.09 excitons per QD. We achieve record low single wavelength gain thresholds down to ~$10^{-5}$ excitons per QD, gain coefficients up to 800 cm$^{-1}$, and a gain lifetime of ~1 ns. Furthermore, we are able to model the gain threshold reduction for the electrochemical charging and the resulting gain lifetimes quantitatively. These results pave the way to achieve optically pumped QD lasers operating at low excitation fluences.

METHODS

**Synthesis of CdSe core nanocrystals (NCs).** The CdSe core nanocrystals were synthesized according to a method by Chen et al.[20]. In a 50 mL three-necked flask, 60 mg of CdO, 280 mg octadecylphosphonic acid (ODPA), 3 g trioctylphosphineoxide (TOPO) and a magnetic stirring bean were added. This mixture of powders was heated up under vacuum to 150°C, where the mixture melts. The mixture was slowly stirred (it prevents the CdO from creeping up the inside of the flask) and degassed at this temperature for one hour. The mixture was heated up to 320 °C, where the liquid turned into a clear and colorless solution. Note that depending on the batch of QDs the time it took for the solution to become clear varied from 20 minutes to 4 hours; this has likely something to do with the impurities in one of the chemicals. 1 mL of trioctylphosphine (TOP) was added to the solution, and the temperature was raised to 380 °C, at which point 0.5 mL of a Se-precursor solution (60mg Se in 0.5mL TOP) was swiftly injected. After a specific growth time the reaction mixture was cooled with an air-gun to room temperature. For the CdSe cores in this work, we used a growth time of ±25 seconds. The crude product is washed once by addition of a 1:1 volume ratio of methyl acetate, followed by centrifugation at 3000 rpm, and redispersion into hexane. Solution is then filtered through several milipore filters (the polymerized ligands clog the filters easily) with a pore diameter of 0.2 μm. The filtered solution is washed and centrifuged again as described above, redispersed in hexane and the resulting sample is stored in a nitrogen purged glovebox for further use.

**Synthesis of Cd-oleate and Zn-oleate for CdS and ZnS shell growth.** For the Cd-oleate synthesis, 1.32 g of Cd-(acetate)2 was dissolved in 52.4 g ODE and 7.4 g OA. The mixture was heated up under vacuum to 120°C and left there for three hours. Afterwards, the reaction was cooled down to room temperature and the Cd-oleate solution was stored in a nitrogen purged glovebox for further use.

The Zn-oleate was made in a similar fashion. $Zn(II)-(acetate)_2$ was mixed with 1g of OA, 1.6 mL ODE and 1.6 mL of OLAM. The oleylamine serves as a stabilizing ligand for the Zn-oleate, since this has the tendency to solidify out of solution at room temperature otherwise. The mixture was heated up in a 20 mL vial inside a nitrogen purged glovebox to 130°C and stored there for further use. Note that the Zn-oleate solution is extremely viscous and should be handled with care when placed into a syringe.

**Shell growth of CdS and ZnS.** The shellgrowth of CdSe QDs into core-shell-shell CdSe/CdS/ZnS nanocrystals was done according to an adapted method by Chen et al.[20], Boldt et al.[21] and Hanafi et al.[19].

For the CdS shell growth, 50 nmol CdSe cores, 3.0 mL octadecene (ODE) and **NO** oleylamine (OLAM, after recent work by Hanafi et al.[19]) were added to a 100 mL three-necked flask and degassed for one hour at room-temperature (21°C) and for 20 hours at 120°C to completely remove hexane, oxygen and water. After that, the reaction solution was heated up to 310°C under nitrogen flow and magnetic stirring. During the heating, when the temperature reached 240°C, a desired amount of Cd-oleate (diluted in ODE) and 1-octanethiol (diluted in 8 mL ODE) were injected dropwise into the growth solution at a rate of half a CdS monolayer per hour using a syringe pump. We define one CdS monolayer as one full layer of Cd and one full layer of S on the NC surface (i.e. half a unit cell). After the addition of the CdS shell-precursors was finished, but before the growth of the ZnS shell, the core-shell QDs containing solution was degassed at a pressure of 0.5 mbar for one hour at 120°C.

For the ZnS shell-growth, the sulfur precursor consisted again out of 1-octanethiol diluted in ODE. The solution with freshly grown CdSe/CdS QDs was heated up to 280°C under nitrogen flow. When the solution reached 210°C, a desired amount of Zn-oleate and 1-octanethiol in 4 mL ODE (in two separate syringes) was injected at a rate of 2 mL/hour (roughly one monolayers of ZnS per hour). After addition of the precursors, the solution was cooled down to room temperature by removing the heat and with an air-gun.

The solution was washed twice by addition of methanol:butanol (1:2), centrifugation at 3000 rpm for 10 minutes, and once with methylacetate followed by centrifugation at 3000 rpm. The precipitate was each time redispersed in hexane. Finally, the solution was filtered through milipore filters with a pore diameter of 0.2 μm and stored in a nitrogen purged glovebox for further use.

Using the above method, we synthesized a batch of core-shell-shell CdSe/8CdS/2ZnS QDs.

**QDs-on-ITO film preparation.** We prepared a concentrated solution (roughly 20 mg/mL) of QDs in toluene. Before spincoating, the ITO slide is cleaned by sonication in isopropanol and rinsing with ethanol and acetone, followed by drying with an airgun. The slide is placed inside a UV-Ozone cleaner for 30 minutes prior to spincoating, to increase the wetting of the QD solution on the ITO. The spincoating was performed by gently dropcasting 40 μL of the QD dispersion on the ITO slide, followed by spincoating for 1 minute at 1000 rpm (with a ramp rate of 200 rpm/s). The film is taken inside a N2 purged glovebox, where we dropcast a solution of 0.5 M 1,7-diaminoheptane in methanol on top of the ITO slide, letting the methanol of this solution evaporate, followed by submerging the substrate into clean methanol. This ligand exchange/stripping procedure is repeated two more times, to ensure proper ligand exchange/stripping. Without performing this treatment, we are not able to electrochemically inject any electrons into the 1S(e) conduction band state of the QD film, as the film is not conductive enough and the electrons cannot hop from QD to QD.

**Steady state absorption and photoluminescence measurements.** Absorption spectra were measured on a double-beam PerkinElmer Lambda 1050 UV/Vis spectrometer; in case of the QD films on ITO, the sample was measured inside an integrating sphere and an empty ITO was measured separately for background correction. Photoluminescence spectra were recorded on an Edinburgh Instruments FLS980 spectrofluorimeter equipped with double grating monochromators for both excitation and emission paths and a 450 W Xenon lamp as an excitation source.

**Transmission Electron Microscopy (TEM).** TEM images were acquired using a JEOL JEM-1400 plus TEM microscope operating at 120 kV. Samples for TEM imaging were prepared by dropcasting a dilute solution of QDs onto a Formvar and carbon coated copper (400-mesh) TEM grid.

**fs-Transient Absorption (TA) spectroscopy.** fs-TA measurements are performed on solutions of the CdSe(/CdS/ZnS) QDs in hexane or toluene, loaded inside an air-tight cuvet inside a nitrogen purged glovebox. A Yb-KGW oscillator (Light Conversion, Pharos SP) is used to produce 180 fs photon pulses with a wavelength of 1028 nm and at a frequency of 5 kHz. The pump beam is obtained by sending the fundamental beam through an Optical Parametric Amplifier (OPA) equipped with a second harmonic module (Light Conversion, Orpheus), performing non-linear frequency mixing and producing an output beam whose wavelength can be tuned in the 310-1330 nm window. A small fraction of the fundamental beam power is used to produce a broadband probe spectrum (480-1600 nm), by supercontinuum generation in a sapphire crystal. The pump beam is transmitted through a mechanical chopper operating at 2.5 kHz, allowing one in every two pump pulses to be transmitted. Pump and probe beam overlap at the sample position with a small angle (roughly 8°), and with a relative time delay controlled by an automated delay-stage. After transmission through the sample, the pump beam is dumped while the probe is collected at a detector (Ultrafast Systems, Helios). During the experiments, we make sure the pump and probe beam have orthogonal polarizations (i.e. one of them is vertically polarized, the other horizontally), to reduce the influence of pump scattering into our detector. The differential absorbance is obtained via $\Delta A = ln(I_{on}/I_{off})$, where $I$ is the probe light incident on the detector with either pump on or pump off. TA data are corrected for probe-chirp via a polynomial correction to the coherent artifact. Pump photon fluence was estimated by measuring the pump beam transmission through a 1-mm-radius pinhole with a thermopile sensor (Coherent, PS19Q).

We also measure transient reflection (TR) spectra to obtain the true change in absorption in transient transmission experiments. The correction method is outlined in the SI.

**Photoluminescence quantum yield (PLQY) measurements.** We measured the PLQY of the NC dispersions with respect to a Rhodamine 101 solution in ethanol. The PLQY was calculated using the following equation;

$$PLQY = PLQY_{Rhodamine\ 101}\ \frac{I^{PL}_{QD\ solution}}{I^{PL}_{Rhodamine\ 101}} \frac{f_{Rhodamine\ 101}}{f_{QD\ solution}} \left(\frac{n_{hexane}}{n_{ethanol}}\right)^2$$

Where PLQY$_{rhodamine\ 101}$ is set to be 95%, I$^{PL}$ is the intensity of the photoluminescence signal of either the QD solution or the Rhodamine 101 solution, n$_{hexane/ethanol}$ is the refractive index of hexane or ethanol at 530 nm (1.377 and 1.3630) and f$_x$ is the fraction of absorbed light of species x, calculated as $f_x = 1 -$ $10^{-OD_x}$, where $OD_x$ is the optical density of the solution containing either the QDs or the Rhodamine 101. We determined the PLQY of the CdSe/8CdS/2ZnS core-shell-shell QDs to be 81%.

**Spectroelectrochemical (SEC) measurements.** The SEC measurements were all performed in a $N_2$ purged glovebox. As an electrolyte, we used an 0.1 M $LiClO_4$ solution in acetonitrile, which was dried with an Innovative Technology PureSolv Micro column. The QD film was immersed in the electrolyte solution, together with a Ag wire pseudoreference electrode and a Pt sheet counter electrode. The potential of the NC film on ITO was controlled with a PGSTAT128N Autolab potentiostat. Changes in the absorption or PL of the NC film as a function of applied potential were recorded simultaneously with a cyclic voltammogram with a fiber-based UV-VIS spectrometer (USB2000, Ocean Optics). For the film, the measurements were started at the open-circuit potential ($V_{OC}$ = -0.3V w.r.t. Ag wire, i.e. -0.77V vs. $Fc/Fc^+$, see SI)), while scanning with a rate of 20 mV/s. Unless stated otherwise, all potentials are given w.r.t. the Ag pseudoreference. For SEC measurements combined with fsTA, we loaded the samples inside a nitrogen purged glovebox into a leak-tight sample holder (see SI for more information).

**Supporting Information available.** Description of the synthesis of the NCs employed in this work. Description of the techniques used to characterize the NCs (TEM, steady-state absorption, PL). Description of the setups used for the TA and spectroelectrochemical measurements. Data analysis: all acquired data, derivation of and the correction for changes in reflection upon photoexcitation, simulating the gain threshold and gain lifetime using a Fermi-Dirac type model.

**Competing financial interests.** The authors declare no competing financial interests.

**Corresponding authors.**   Arjan J. Houtepen: a.j.houtepen@tudelft.nl

Jaco J. Geuchies: j.j.geuchies@tudelft.nl

**Acknowledgements.** AJH, JJG, SG and WvdS gratefully acknowledge financial support from the European Research Council Horizon 2020 ERC Grant Agreement No. 678004 (Doping on Demand). GG acknowledges financial support from NWO-TTW (Project No. 13903, Stable and Non-Toxic Nanocrystal Solar Cells). We gratefully acknowledge fruitful discussions with Dr. Freddy Rabouw, Stijn Hinterding and Sander Vonk (Utrecht University) on modelling the electrochemical electron injection into QDs.

**Data availability.** The data and that supports the finding of this study and the codes used to analyze them (Python) are available upon reasonable request from the corresponding authors upon reasonable request, and will be uploaded on a repository shortly.

# Supporting Information

# Quantitative electrochemical control over optical gain in quantum-dot solids

Jaco J. Geuchies[†], Baldur Brynjarsson[†], Gianluca Grimaldi[†*], Solrun Gudjonsdottir[†], Ward van der Stam[†$], Wiel H. Evers[†], Arjan J. Houtepen[†]

[†] Optoelectronic Materials Section, Faculty of Applied Sciences, Delft University of Technology, Van der Maasweg 9, 2629 HAZ Delft, The Netherlands

[*] Current address: Center for Nanophotonics, AMOLF, Science Park 104, 1098 XG Amsterdam, The Netherlands

[$] Current address: Inorganic Chemistry and Catalysis, Debye Institute for Nanomaterials Science, Utrecht University, Universiteitsweg 99, 3584 CG Utrecht, The Netherlands.

# Table of Contents - Supporting Information.

# METHODS

**Materials:** Lithium perchlorate (LiClO$_4$, 99.99%), CdO (99.99%), Cd(II)-acetate (99.995%), Zn(II)-acetate (99.99%), Rhodamine 101 inner salt, Octadecene (ODE, 90%), 1,7-heptanediamine (7-DA, 98%), Oleylamine (OLAM, 99.8%), Oleic acid (OA, 90%), Ferrocene (Fc, 98%), Butanol (BuOH, Anhydrous, 99.8%), Methanol (MeOH, Anhydrous, 99.8%), Hexane (99.8%, Anhydrous), Octadecylphosphonic acid (ODPA, 97%), Trioctyl phosphineoxide (TOPO, technical grade, 90%), Trioctylphosphine (TOP, 97%), 1-Octanethiol (>98.5%), Selenium powder (Se, 99.99), and Acetonitril (99.99%, Anhydrous) were all bought from Sigma-Aldrich and used as received unless specifically mentioned. Acetonitril was dried before use in an Innovative Technology PureSolv Micro column. All other chemicals were used as received, unless specifically mentioned.

We measure transient reflection (TR) spectra to obtain the true change in absorption in transient transmission experiments. The correction method is outlined in a later section of the SI.

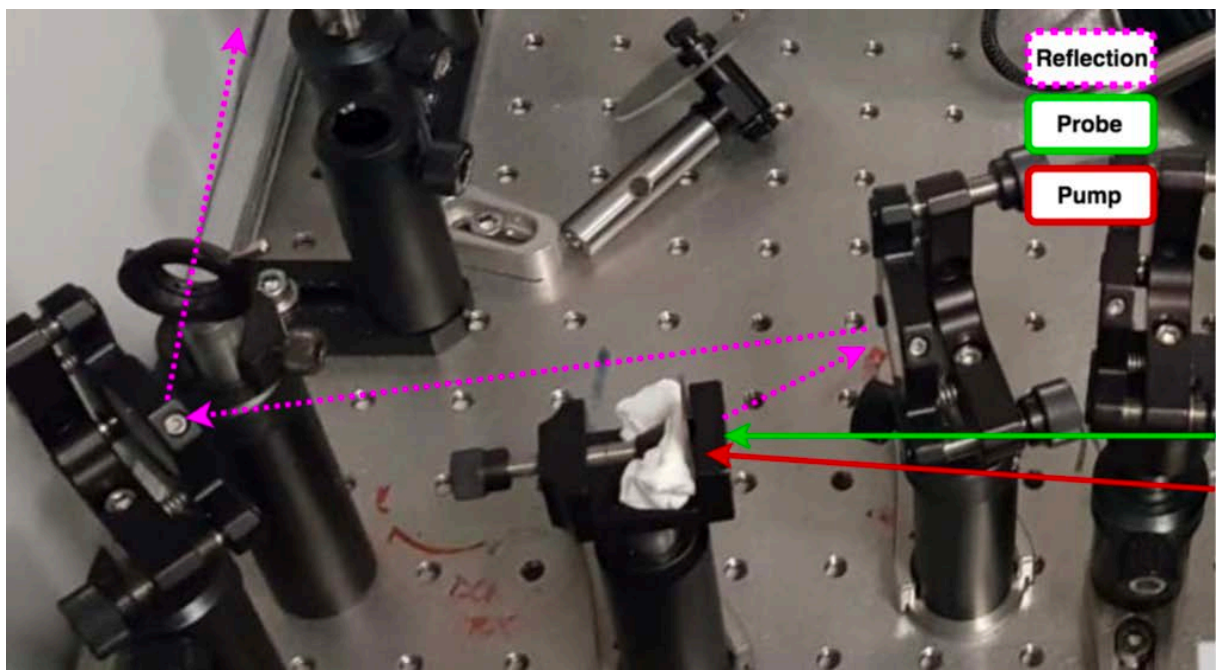

**Figure S1: Alignment of pump and probe for the transient reflection experiments.** The transmitted probe beam is blocked after the sample (not shown in picture). Also here, pump and probe polarizations are orthogonal to each other.

# METHODS

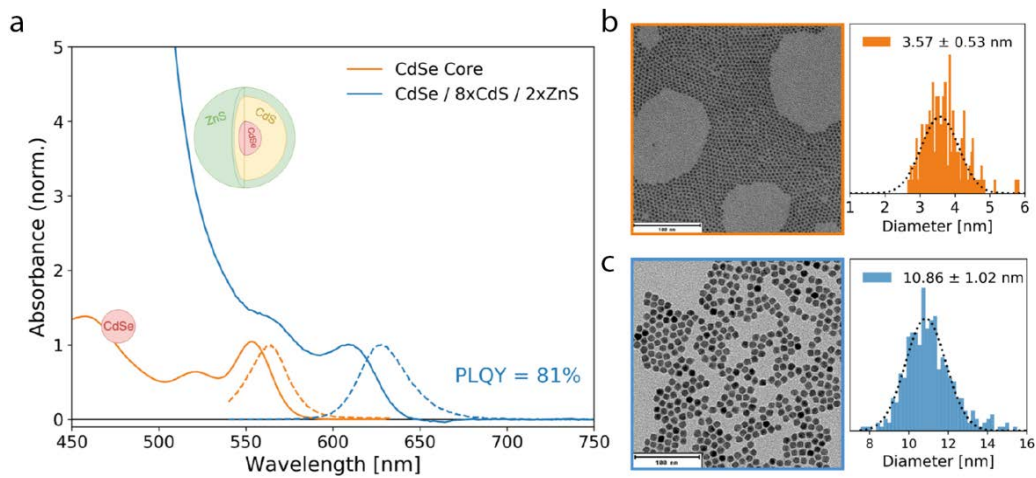

**Figure S2: Optical and structural characterization of the QDs used throughout this study.** **(a)** Absorption and PL of the core CdSe QDs (orange) and core-shell-shell (blue) QDs. The quantum yield of the final cores is measured to be 81% w.r.t. a reference dye. **(b)** Representative TEM image of the cores, with a histogram of the measured sizes as analyzed with standard ImageJ routines. **(c)** Same as in **(b)**, but for the core-shell-shell QDs.

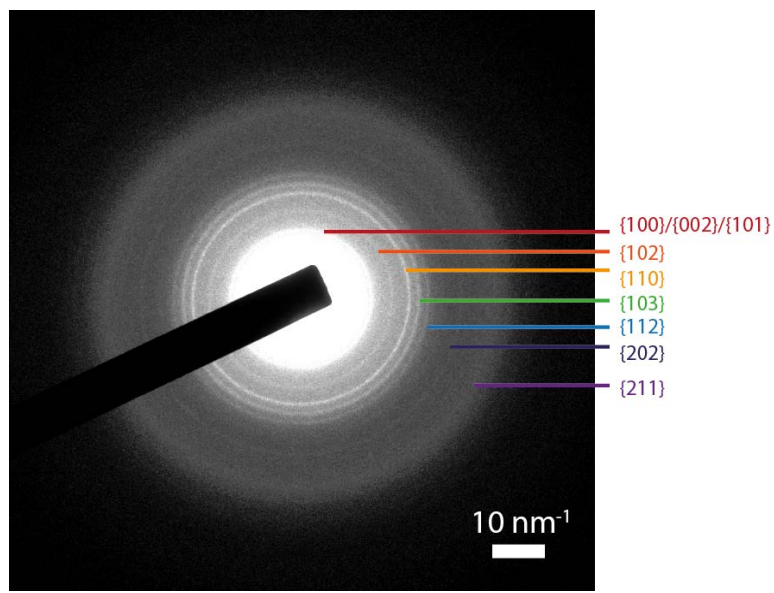

**Figure S3: Electron diffraction of the core-shell-shell QDs.** Reflections are indexed according to wurtzite CdSe/CdS (see ICDD JCPDS: CdSe 00-019-0191). The first triplet of peaks is not very well resolved due to saturated camera intensity. The second triplet of diffraction peaks (originating from the {110}, {103} and {112} wurtzite lattice planes) can be clearly distinguished.

# Supplementary discussion 1: Modelling the optical gain threshold

## 1.1 Absorption and gain threshold for neutral QDs – transition counting

In this model we do not regard spectral shifts of the photoexcited QDs, and we take the oscillator strength of the band-edge transition to be proportional to the number of available combinations of an unoccupied valence and conduction band-edge states. In the ground state, this oscillator strength of the band-edge (1S) transition for a single QD can be written as;

$$A_0 = \gamma g_e g_h \qquad (\text{eq. S1})$$

Where $g_e$ and $g_h$ are the degeneracies of the 1S(e) conduction band and $1S_{3/2}$(h) valence band levels respectively, and $\gamma$ is some proportionality constant. In the excited state, some absorption transitions are blocked by photoexcited excitons, which we write down as

$$A^*(n_e, n_h) = \gamma [(g_e - n_e)(g_h - n_h) - n_e n_h] \qquad (\text{eq. S2})$$

Where $n_e$ is the number of electrons in the 1S(e) state and $n_h$ the number of holes in the $1S_{3/2}$(h) state per QD. The left-hand-side of equation S2 corresponds to the number of absorption transitions that are possible, whereas the right-hand-side corresponds to the number of stimulated emission transitions that are possible. Upon photoexcitation, $n_e = n_h = N$, the number of excitons in a QD, and equation S2 simplifies to

$$A^*(N) = \gamma [(g_e - N)(g_h - N) - N^2] \qquad (\text{eq. S3})$$

When a number of electrons in the 1S(e) state larger than its degeneracy (two) is present, it should not contribute to the band-edge absorption and stimulated emission anymore. Equation S3 (and S2 for that matter) can be adjusted to take this into account;

$$A^*(N) = \gamma(Max[(g_e - N), 0]Max[(g_h - N), 0] - Min[N, g_e]Min[N, g_h]) \quad \text{(eq. S4)}$$

The Max[a,b] (and Min) functions take the maximum (minimum) value of a or b. For example, when N = 1 the first function Max[2 - 1, 0] = 1. When N = 2, the first function Max[2 - 2, 0] = 0. Now when the number of electrons becomes larger than the degeneracy $g_e$ = 2, for example N = 3, then the first function Max[2 - 3, 0] = 0; the third electron does not cause a 'negative' absorption term, rather it will not contribute to the band-edge transition's oscillator strength. There are subtle differences between equations S3 and S4 after going from a single particle picture to an ensemble picture. Both equations S3 and S4 are plotted in Figure S4(a). Equation S4 can also not be solved analytically anymore to obtain a threshold (where A* becomes zero). Again, the above equations in practice holds for a single QD. We use $g_e$ = 2 and $g_h$ = 4 unless mentioned otherwise.

In an ensemble of QDs, e.g. in solution or in a film, we create a Poisson distribution of excitations, and generate a certain average number of excitons per QD, <N>. For each <N> we can calculate the Poisson distribution, and calculate the absorption for each N, and sum them up weighted by their relative Poissonian probability;

$$A^*(\langle N \rangle) = \gamma \sum_{N=0}^{i} \frac{\langle N \rangle^N}{N!} e^{-\langle N \rangle} A^*(N) \quad \text{(eq. S5)}$$

Note that <N> and N are separate variables. The summation runs over all populations that have an N number of excitons. Care needs to be taken into account that the summation over N runs up to higher values than one wants to quantify for <N> (in our case we run our summation up to i = 20, and our experimental data goes up to <N> = 9). In reality, one can also omit N's for which the Poissonian weight becomes less than 1%. The gain threshold, the average number of excitons needed per QD in order to set A*(<N>) = 0, can be obtained numerically, and equals

<N_gain> = 1.54. This gain threshold should hold for the isolated QDs in solution and for the neutral QD film (i.e. at the open circuit potential).

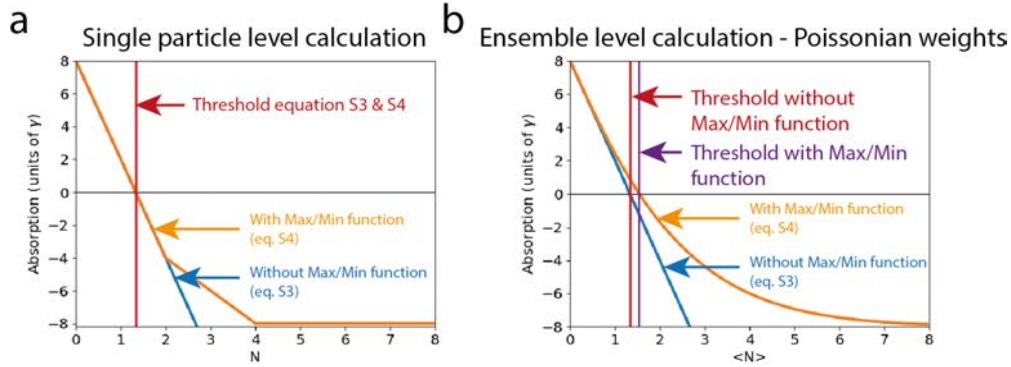

**Figure S4:** Band-edge absorption as a function of exciton population per QD on **(a)** single particle level and **(b)** ensemble level, i.e. eq. S3 and eq. S4 plugged into eq. S5. Note the subtle difference in optical gain threshold with and without properly accounting for saturation of the transitions (4/3 ≈ 1.33 excitons/QD not taking this into account [eq. S3] vs. 1.54 excitons/QD when properly taken into account [eq. S4]).

## 1.2 Absorption and gain threshold for charged QDs

The absorption for QDs charged with additional electrons can be calculated by adjusting equation S2/S3. We omit noting γ from now on and will always calculate the absorption in units of γ;

$$A^*(n_e, n_h) = [(g_e - n_e)(g_h - N) - n_e N] \qquad \text{(eq. S5)}$$

With $n_e = N + n_{echem}$, the total number of electrons in the 1S(e) state being equal to the number of excitons optically excited in the QD plus the number of electrochemically injected electrons into this state. Using the same formalism as above, we take into account that when more electrons (or holes) than the degeneracy of the band-edge states are present, they should not contribute to absorption and stimulated emission terms of the band-edge transition;

$$A^*(N, n_{echem}) = Max(g_h - N, 0)Max(g_e - N - n_{echem}, 0) - Min(N, g_h)Min(N + n_{echem}, g_e). \quad \text{(eq. S6)}$$

Equation S6 is equal to equation 1 in the main text. This equation assumes that there is no distribution in doping levels, i.e. every QD in the ensemble has exactly $n_{echem}$ electrons in their 1S(e) level. The results of equation S6 (on an ensemble level, i.e. plugged into equation S6, are plotted in Figure 1(g) of the main text). The optical gain threshold is reduced to zero excitons/QD by charging the 1S(e) level with 2 electrons per QD.

## 1.3 Fermi-Dirac distribution for electron filling into the 1S(e) state

In the discussion above we assumed that there was no distribution of electron occupancies. A more realistic approach would be to assume a Fermi-Dirac distribution of the electrons throughout the 1S(e) state;

$$n_i = \frac{1}{e^{\frac{\varepsilon_i - \mu}{k_B T}} + 1} \quad \text{(eq. S6)}$$

Where $n_i$ will give the fractional occupancy of state i, $\varepsilon_i$ is the energy of state i, $\mu$ is the chemical potential, $k_B$ is Boltzmann's constant, $T$ temperature. To get the total number of electrons in state i, we can multiply $n_i$ with the degeneracy of the state ($g_e$ = 2). However, we obtain information on a macroscopic scale, not on the microscopic occupancy of the state, i.e. we cannot obtain the fraction of neutral, singly charged and doubly charged through this formalism. These can be obtained as follows; we start by calculating the grand canonical partition function, in which the system can exchange both energy and particles, which has the general form:

$$\mathcal{Z} = \sum_i g_i e^{\beta(N_i\mu - \varepsilon_i)} \qquad \text{(eq. S7)}$$

Where the sum runs over all microstates where $N_i$ particles have energy $\varepsilon_i$ in contact with a heat reservoir with $\beta = 1/k_B T$, and $g_i$ the number of microstates with the same energy $\varepsilon_i$. We rewrite the equation in units of potential instead of chemical potential ($V = \mu/e$, with $e$ the elementary charge) and the energy of state i as a potential difference ($\varepsilon_i = eV_i$);

$$\mathcal{Z} = \sum_i g_i e^{\beta e(N_i V - V_i)} \qquad \text{(eq. S8)}$$

We now write the partition function for all three microstates; neutral (I), singly charged (II) and doubly charged (III) species:

I. Neutral quantum dots. We inject zero electrons, so $N_0 = 0$ and we set $V_0 = 0$. The degeneracy of this microstate $g_0 = 1$ and partition function becomes $\mathcal{Z}_0 = 1$.

II. Singly charged QDs. Since we inject one electron, $N_1 = 1$, and we set $V_1 = V_{1S}$, the potential of the 1S(e) electron state. Since we can fill either state from the twofold degenerate 1S(e), the degeneracy of this microstate $g_1 = 2$. The partition function for this microstate becomes $\mathcal{Z}_1 = 2e^{\beta e(V - V_{1S})}$.

III. Doubly charged QDs. We completely fill the 1S(e) level with electrons; $N_2 = 2$, and $V_2 = 2V_{1S}$. There might be some charging energy $V_C$ related to injection of the second electron into the 1S(e) state, which we can take into account as $V_2 = 2V_{1S} + V_C$. The degeneracy for this microstate $g_2 = 1$, so the total partition function becomes $\mathcal{Z}_2 = e^{\beta e(2V - 2V_{1S} - V_C)}$

The total partition function for our 1S(e) level is equal to the sum of I, II and III and becomes

$$\mathcal{Z} = 1 + 2e^{\beta e(V-V_{1S})} + e^{\beta e(2V-2V_{1S}-V_C)} \qquad \text{(eq. S9)}$$

The fraction of neutral $f_0$, singly charged $f_1$ and doubly charged $f_2$ QDs as a function of applied potential can be now calculated using

$$f_i = \frac{1}{\mathcal{Z}} g_i e^{\beta e(N_i V - V_i)} \qquad \text{(eq. S10)}$$

and we obtain for the three fractions (as shown in I, II and III)

$$f_0 = \frac{1}{\mathcal{Z}}, \quad f_1 = \frac{1}{\mathcal{Z}} 2e^{\beta e(V-V_{1S})}, \quad f_2 = \frac{1}{\mathcal{Z}} e^{\beta e(2V-2V_{1S}-V_C)} \qquad \text{(eq. S11-S13)}$$

The average number of electrons at each potential can be written based on these fractions;

$$\langle n_e \rangle (V_{1S}, V_C, \beta) = 0 \cdot f_0 + 1 \cdot f_1 + 2 \cdot f_2 \qquad \text{(eq. S14)}$$

We fitted equation S14 to the data showing <$n_e$> versus potential to the data presented in Figure 5b of the main text. For a given temperature, equation S14 equals the Fermi-Dirac distribution (equation S6, multiplied with 2, the degeneracy of the 1S(e) level) which is plotted in Figure S5(a).

One could $\beta$ as a fit parameter for the potential vs. <$n_e$> data, where the effective temperature will act as measure for the disorder, and hence the width of the graph. We think the more physical approach is to fix $\beta$ (to room temperature, which is $\approx$ 40 eV$^{-1}$) and instead write the 1S(e) state as a broadened normal distribution;

$$V_{1S}(V) = \frac{1}{\sigma_{1S}\sqrt{2\pi}} e^{\frac{(V-V_{1S})}{2\sigma_{1S}^2}} \qquad \text{(eq. S15)}$$

Where $\sigma_{1S}$ now contains the width of the curve. We could fix $\sigma_{1S}$ to the width of the absorption peak of the 1S transition ($\approx$ 50 meV), but we choose to keep it as a fit parameter. Next to the linewidth of the 1S absorption transition, it now also contains other sources of disorder in the system (e.g. local variations of (dielectric) environment around QDs, electrostatic disorder, etc.). We now have to integrate equation S14 over the potential;

$$\langle n_e \rangle(V_{1S}, V_C, \sigma_{1S}) = \frac{1}{\sigma_{1S}\sqrt{2\pi}} \int e^{\frac{(V-V_{1S})}{2\sigma_{1S}^2}} (1 \cdot f_1(V, V_{1S}) + 2 \cdot f_2(V, V_{1S}, V_C)) \, dV \qquad \text{(eq. S16)}$$

We fit a numerical form of this integral and use $V_{1S}$, $V_C$ and $\sigma_{1S}$ as fit parameters. When either equation S14 or equation S16 are fitted, we obtain the same result, as shown in Figure S5(b).

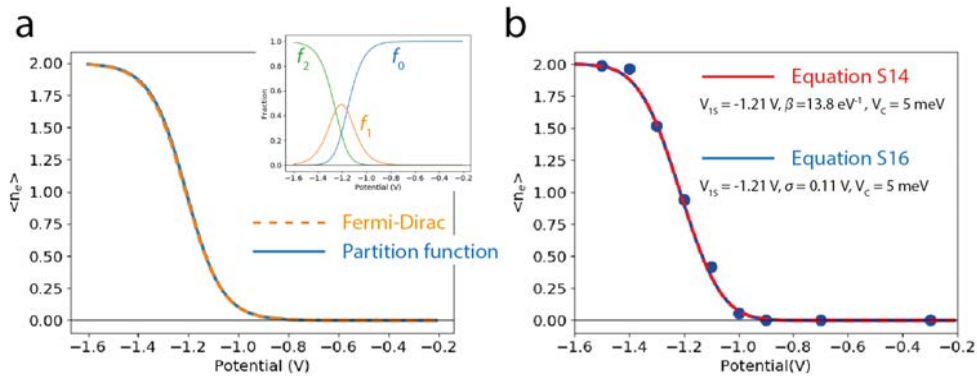

**Figure S5:** Fermi-Dirac statistics for the electron injection process into the 1S(e) level. **(a)** Overlap of equation S6, The Fermi-Dirac Function, with equation S14, the partition-function-based approach. They are identical. The inset shows the populations extracted from the fit to the electrochemical data. **(b)** Comparison between equation S14, using an effective $\beta$ (in other words temperature) as a fit parameter to account for disorder and broadening, with equation S16, where we use the width of the normal distribution of the 1S(e) electron level $\sigma_{1S}$ as a fit parameter for disorder and broadening.

When using $\beta$ as fit parameter, we obtain an effective temperature of the system of 841K. Instead, now using the width of the normally distributed 1S level, we obtain a width of $\sigma_{1S}$ = 110 mV. This is significantly broader than the width of the 1S absorption transition ($\approx$ 50 meV); we hypothesize that this additional broadening is due to electrostatic (and other forms of) disorder.

We varied the charging energy $V_C$ that we obtain from the fit (5 meV) and plot the results in Figure S6. As long as the charging energy is small compared to the width from the 1S potential $V_{1S}$, i.e. $V_C < \sigma_{1S}$, the charging energy does not impact the resulting fit by much. We also plot the predicted optical gain threshold from the film when we assume there is no distribution of $n_e$, i.e. each QD gets charged with exactly the same number of electrons, which is plotted in Figure S6(b). The resulting modelled curve close matches the experimental data, however we think that this is not a reasonable description of the charging process and that the Fermi-Dirac distribution of electrons throughout the film is physically more correct.

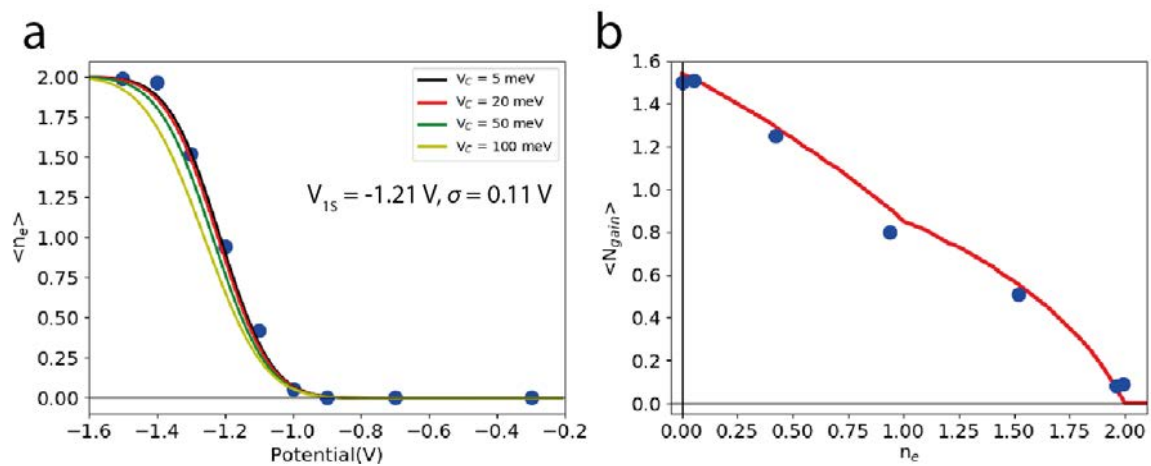

**Figure S6: (a)** Influence of the charging energy $V_C$ for the injection of the second electron on the resulting fit; as long as it is significantly smaller than the width of the distribution, it does not impact the resulting fit much. Datapoints are show as blue dots and the predicted fits are shown as solid lines. **(b)** Modelled gain threshold $<N_{gain}>$ versus the number of electrons per

QD $n_e$ (red line), where we now assume there is no distribution of $n_e$ over the QD film. Applying a Fermi-Dirac distributions smooths out the kink in the model (see Figure 5a of the main text).

## 1.4 Obtaining a gain threshold for each doping density $\langle n_e \rangle$

We now outline the model for acquiring an optical gain threshold for each doping density. The workflow of the model is schematically illustrated in Figure S7.

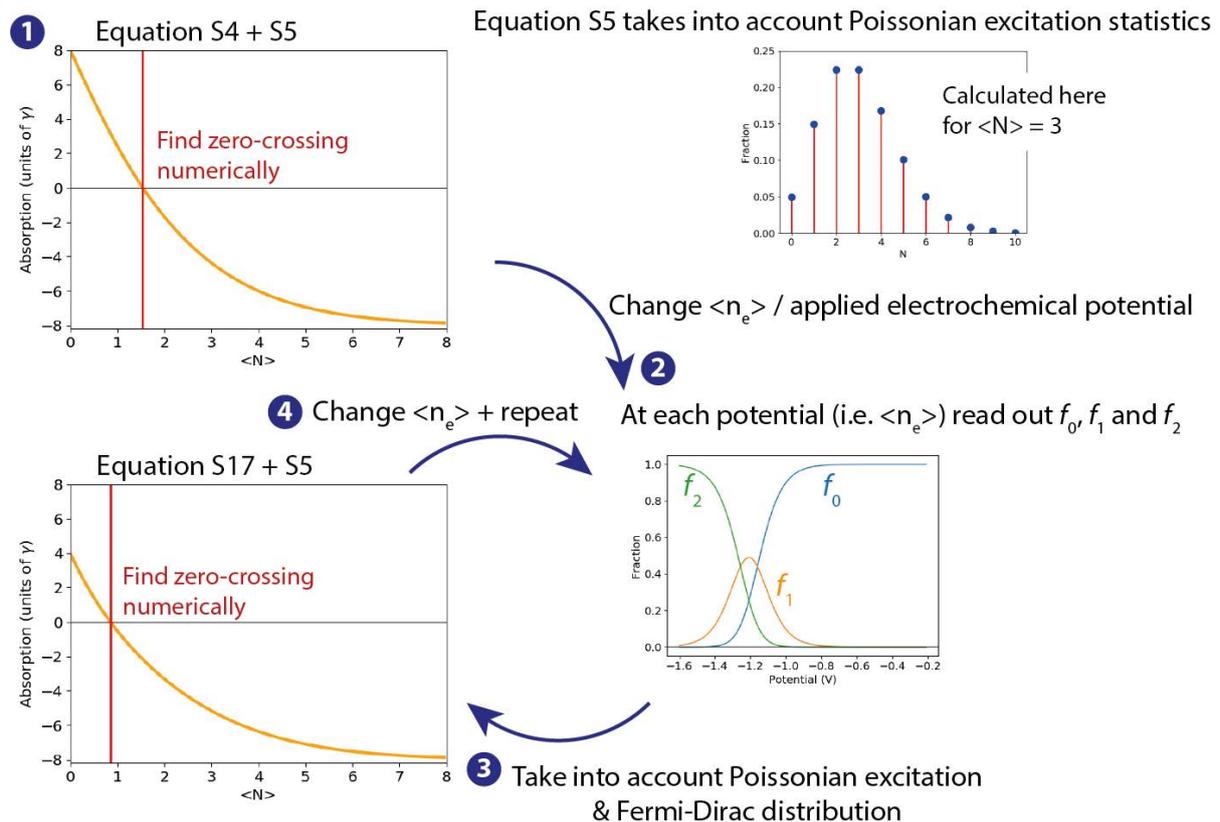

**Figure S7: workflow for modelling the optical gain threshold as a function of the average number of electrochemically injected electrons.**

From equation S6, we can calculate the absorption amplitude of the band-edge transition for QDs with a certain doping level (zero, one or two electrons). We use the obtained the fraction of neutral ($f_0$), singly-charged ($f_1$) and doubly charged ($f_2$), and use these as weighing factors to calculate the absorption at a given *average* doping density $<n_e> = f_0 \cdot 0 + f_1 \cdot 1 + f_2 \cdot 2$

$$A^*(N, n_{echem}, f_0, f_1, f_2) = f_0 \cdot A^*(N, n_{echem} = 0) + f_1 \cdot A^*(N, n_{echem} = 1) + f_2 \cdot A^*(N, n_{echem} = 2) \quad (eq.\ S17)$$

Each term in equation S17 corresponds to equation S6, with a varied number of electrochemically injected electrons into the 1S(e) state $n_{echem} = 0, 1$, or 2. The absorption is calculated according to equation S5 (which takes into account the Poissonian excitation of the QD film), where we plug in equation S17. The excitation density at which the absorption becomes zero, i.e. $<N_{gain}>$, is found numerically, and the process is repeated for different levels of $<n_e>$.

Summarizing, the full equation for computing the absorption as a function of the number of excitons and electrochemically injected electrons into the 1S(e) level can be computed as:

$$A^*_{ensemble}(\langle N_X \rangle, \langle n_{1S(e)} \rangle) = \sum_{n_{echem}=0}^{2} \sum_{N=0}^{N_{max}} f(\langle n_{1S(e)} \rangle) \cdot P(\langle N_X \rangle, N) \cdot A^*(N, \langle n_{1S(e)} \rangle)$$

(eq. S17)

# 1.5 Predicting the gain lifetime based on the biexciton and triexciton Auger rates

We now adapt the models of section 1.3 and section 1.4 to predict the time-dependent evolution of the absorption based on the experimentally obtained Auger rates for the decay of the biexcitons and triexcitons in the film.

After photoexcitation, the average number of excitons per QD, <N>, starts decreasing, as a consequence of Auger recombination. To predict the time-dependence of the excited state absorption we need to compute the initial fraction of the QD population in each multiexciton state, and then compute the evolution in time of those fractions as Auger recombination takes place.

To compute the initial fraction of QDs in each multiexciton state, we first used Poisson statistics to calculate the fraction of QDs photoexcited with $N_i$ excitons given an average number <N> of excitons in the QD ensemble, $f_P(N_i,<N>)$ for $N_i=0,\ldots, N_{max}$, as described in Section 1.1. We then computed the fraction of QDs with 0, 1 and 2 electrochemically injected electrons, $f_{FD}(n_e)$ for $n_e=0,1,2$, as described in Section 1.3. Finally, the fraction of QDs that can be found at t=0 in the multiexciton state characterized by $N_i$ photoexcited exciton and $n_e$ electrochemically injected electrons will be given by:

$$f(N_i, n_e, t = 0) = f_P(N_i, \langle N \rangle) * f_{FD}(n_e) \quad \text{(eq. S18)}$$

In order to evolve the multiexciton population in time, we need to solve a system of differential equations coupling together different multiexcitonic states. Since Auger recombination results in the recombination of one or more of the excitons in a QD, the imbalance between negative and positive charges, set by the initial number of electrochemically injected electrons, is conserved throughout the recombination. Therefore, we can split the system of differential equations in three coupled systems, characterized by the number of electrochemically injected electrons (i.e. multiexcitons with different $n_e$ are not coupled to each other):

$$\frac{df}{dt}(N_{max}, n_e, t) = -k(N_{max}, n_e) * f(N_{max}, n_e, t)$$

$$\frac{df}{dt}(N_{max} - 1, n_e, t)$$

$$= -k(N_{max} - 1, n_e) * f(N_{max} - 1, n_e, t) + k(N_{max}, n_e) * f(N_{max}, n_e, t)$$

$$\frac{df}{dt}(2, n_e, t) = -k(2, n_e) * f(2, n_e, t) + k(3, n_e) * f(3, n_e, t)$$

$$\frac{df}{dt}(1, n_e, t) = k(2, n_e) * f(2, n_e, t) - k(1, n_e) * f(1, n_e, t)$$

$$\frac{df}{dt}(0, n_e, t) = k(1, n_e) * f(1, n_e, t) \qquad \text{(eq. S19)}$$

where $N_{max}$ is the highest number of photoexcited excitons that is taken into account in the model, as explained in Section 1.1, and $k(N, n_e)$ is the Auger decay rate of the multiexciton with $N+n_e$ electrons and $N$ holes. In the model, the recombination rate of the single exciton, $k(1,0)$, is set to zero, corresponding to a negligible amount of radiative recombination and trapping during the time-range in which Auger recombination occurs. The recombination rate of the biexciton, $k(2,0)$, and of the triexciton, $k(3,0)$, is obtained from experimental transient absorption data shown in Figure S27. All other recombination rates, which are more difficult to obtain experimentally, can be estimated from the biexciton rate. The Auger recombination rate of the biexciton can be expressed as the product as the rate of a single Auger transition and the number of Auger transition that can take place. Figure S8 shows the eight different Auger transitions allowed in a biexciton. Dividing the biexciton recombination rate by eight gives the rate of a single Auger transition, $k_{Aug}$. The rate of all other multiexciton transitions can then be estimated multiplying $k_{Aug}$ by the number of Auger transition allowed in the multiexciton state.

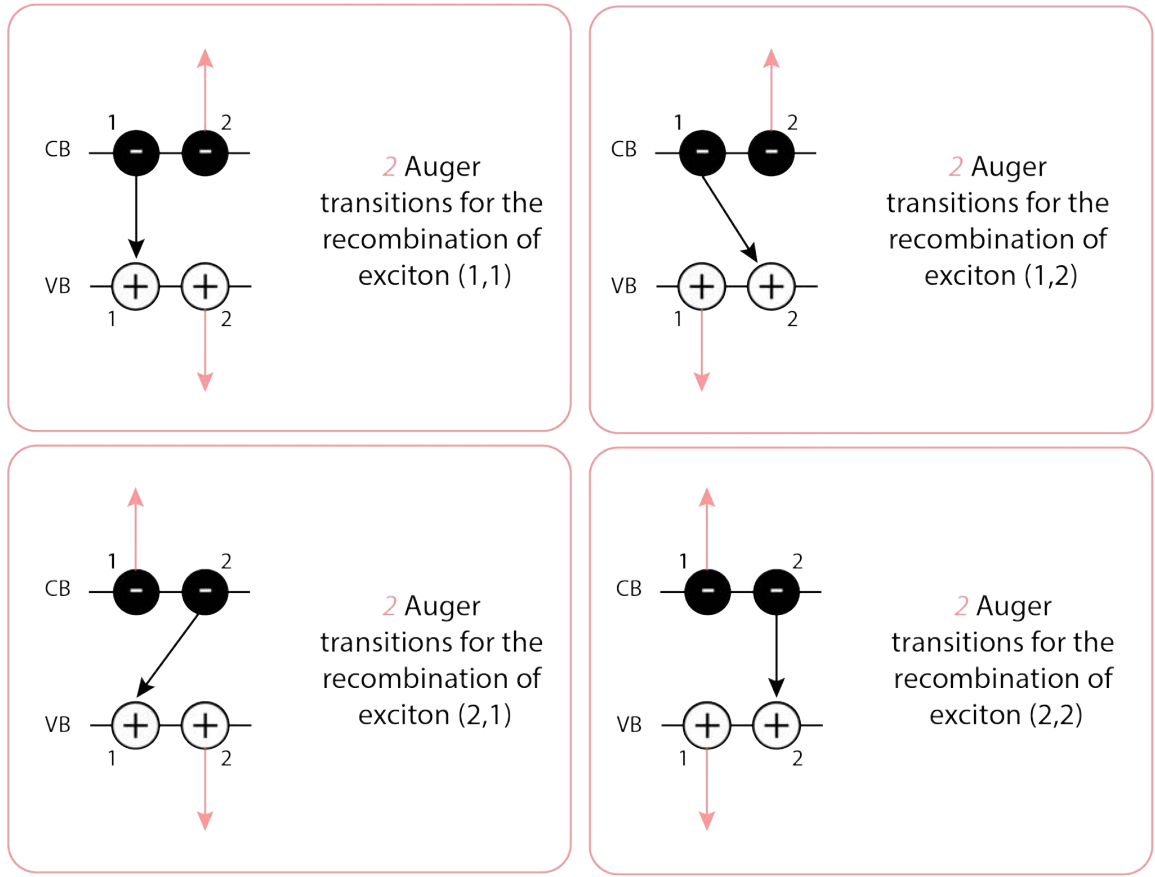

Total number of biexciton Auger transitions: 8

**Figure S8: scheme representing all the Auger transitions available in a biexciton.** Each square highlights the Auger transitions originating from the recombination of each electron-hole pair (black arrow), showing the two possible ways the energy of the pair can be given to one of the remaining carriers (salmon arrow). The total number of Auger transitions for the biexciton is 8, and the rate of Auger recombination of other multiexcitons can be estimated scaling the biexciton recombination rate by the number of Auger transitions available in the multiexciton.

If a multiexciton has *e* electrons and *h* holes, the number of electron-hole pair that can recombine via Auger recombination is *e*h*, while the number of remaining carriers that can accept the energy of the recombining pair is max(*e*-1,0)+max(*h*-1,0). Therefore, the total number of Auger transitions is:

$$N_{Aug} = e * h * (\max(e - 1,0) + \max(h - 1,0)) \qquad (eq.\ S20)$$

In our experiment, all holes are photogenerated, while a certain number of electrons (0, 1, or 2) can be injected electrochemically. Therefore, the formula for the number of Auger transitions in a multiexciton with N photoexcited excitons and $n_e$ electrochemically injected electrons is:

$$N_{Aug} = (N + n_e) * N * (\max(N + n_e - 1,0) + \max(N - 1,0)) \qquad (eq.\ S21)$$

Therefore, the recombination rate of the multiexciton will be:

$$k(N, n_e) = k_{Aug} * N_{Aug} = \frac{k(2,0)}{8} * (N + n_e) * N * (\max(N + n_e - 1,0) + \max(N - 1,0))$$

(eq. S22)

Utilizing the values for the Auger recombination rates in eq. S22 and the initial multiexciton distribution set by eq. S18, the set of differential equations in S19 can be solved as a function of time, obtaining the value at time t of the fraction of QDs with N photoexcited excitons and $n_e$ electrochemically injected electrons, f(N,$n_e$,t), for $n_e$=0,1,2 and N=0,..., $N_{max}$. The set of differential equation is integrated numerically in a python script employing the scipy.integrate.odeint function (See SX). The integration is performed from t=0 to t=2.5 ns, taking 6.25 ps time steps (400 time points). Knowing every f(N,$n_e$,t), the excited state absorption of the QD population at time t, $A^*_{ensemble}(N, n_{echem}, t)$, can be calculated with the time-dependent version of eq. S17:

$$A^*{}_{ensemble}(N, n_{echem}, t) = \sum_{n_e=0}^{2} \sum_{N=0}^{N_{max}} f(n_e, N, t) * A^*(N, n_{echem})$$

Finally, the gain lifetime is obtained finding the time $t_G$ for which $A^*{}_{ensemble}(N, n_{echem}, t_G) \sim 0$.

## 1.6 Obtaining the true Transient-Absorption signal: correcting the Transient-Transmission data for changes in reflection of the sample upon photoexcitation

Upon photoexcitation in a fsTA experiment, one records the change in transmission of a broadband probe-pulse as a function of time. This change in transmission is usually directly converted into a ΔA signal; a change in absorption of the sample. However, when there are changes in reflection simultaneously, due to changes in the real part of the dielectric function of the sample, this conversion becomes incorrect. This can be observed clearly as a below-bandgap bleach (reduction in A), which in reality is a change in reflectivity of the sample (shown in Figure SXXa). When measuring a change in reflection and reflectivity spectrum, one can convert the obtained transient 'exctinction' signal into a true change in absorption.

We first define fractions of absorbed, reflected and transmitted light;

$$F_A = \frac{I_A}{I_0} \qquad (eq.\ S23)$$

$$F_R = \frac{I_R}{I_0} \qquad (eq.\ S24)$$

$$F_t = \frac{I_t}{I_0} \qquad (eq.\ S25)$$

Where $F_A$, $F_R$ and $F_t$ are the fractions of absorbed, reflected and transmitted light respectively. $I_0$ is the incoming light intensity, and $I_A$, $I_R$ and $I_t$ stand for reflected, absorbed and transmitted light intensity. As mentioned before, we assume the films do not scatter, and the following relations should hold;

$$F_t + F_A + F_R = 1 \qquad \text{(eq. S26-1)}$$

$$I_t + I_A + I_R = I_0 \qquad \text{(eq. S26-2)}$$

When the film has a reflectance of 0, i.e. $F_R = 0$, we can write that $F_t = 1 - F_A$ and

$$A = -{}^{10}\log(1 - F_A) = -{}^{10}\log(F_t) \qquad \text{(eq. S27)}$$

or $F_A = 1 - F_t = 1 - 10^{-A}$. When a substrate has a finite reflectance, this is no longer valid, since not all the light that reaches the detector (the transmittance) is lost due to absorption;

$$F_R + F_t = 10^{-A} = 1 - F_A \qquad \text{(eq. S28)}$$

Note that now the fraction of absorbed light only depends on the absorbance (r.h.s. of equation SXX).

The differential absorbance that is regularly computed in a TA experiments ($\Delta A^*$), i.e. from equation S27, and the corresponding differential reflectance ($\Delta R$) are given by;

$$\Delta A^* = -{}^{10}\log\left(\frac{I_{t,on}}{I_{t,off}}\right) \qquad \text{(eq. S29)}$$

$$\Delta R = -^{10}\log\left(\frac{I_{R,on}}{I_{R,off}}\right) \quad \text{(eq. S30)}$$

Where $I_{t,on}$, $I_{t,off}$, $I_{R,on}$ and $I_{R,off}$ are the transmitted probe light with (on) or without (off) and the reflected probe light with (on) or without (off) respectively.

We now compute the absorption, when taking into account reflectance of a substrate, i.e. we plug equations S26-1 and S26-2 into equation S27 to obtain:

$$A = -^{10}\log(1 - F_A) = -^{10}\log(F_R + F_t) = -^{10}\log\left(\frac{I_R + I_t}{I_0}\right) \quad \text{(eq. S31)}$$

The true differential absorption, $\Delta A$, is now given by:

$$\Delta A = A_{on} - A_{on} = -^{10}\log\left(\frac{I_{R,on} + I_{t,on}}{I_0}\right) + ^{10}\log\left(\frac{I_{R,off} + I_{t,off}}{I_0}\right)$$

$$= -^{10}\log\left(\frac{I_{R,on} + I_{t,on}}{I_{R,off} + I_{t,off}}\right)$$

We continue rewriting this expression;

$$\Delta A = -^{10}\log\left(\frac{\frac{I_{R,on}}{I_{R,off}} \cdot I_{R,off}}{I_{R,off} + I_{t,off}} + \frac{\frac{I_{t,on}}{I_{t,off}} \cdot I_{t,off}}{I_{R,off} + I_{t,off}}\right)$$

$$= -^{10}\log\left(\frac{10^{-\Delta R} \cdot I_{R,off}}{I_{R,off} + I_{t,off}} + \frac{10^{-\Delta A^*} \cdot I_{t,off}}{I_{R,off} + I_{t,off}}\right) \quad \text{(eq. S32)}$$

Equation S32 only contains static variables (all containing pump-off intensities), so we should be able to rewrite this in terms of measurable quantities, i.e. fractions of reflected and absorbed light. Multiplying both fractions in the logarithm with $I_0/I_0$ yields

$$\Delta A = -{}^{10}\log \left( \frac{10^{-\Delta R}}{(I_{R,off} + I_{t,off})/I_0} \cdot \frac{I_{R,off}}{I_0} + \frac{10^{-\Delta A^*}}{(I_{R,off} + I_{t,off})/I_0} \frac{I_{t,off}}{I_0} \right)$$

$$= -{}^{10}\log \left( \frac{10^{-\Delta R} F_R}{F_R + F_t} + \frac{10^{-\Delta A^*} F_T}{F_R + F_t} \right)$$

$$= -{}^{10}\log \left( \frac{10^{-\Delta R} F_R + 10^{-\Delta A^*}(1-F_A-F_R)}{1-F_A} \right) \quad \text{(eq. S33)}$$

Where we have rewritten the fraction of transmitted light $F_t$ in to fractions of absorbed and reflected light (as measured in absorption spectrometers with integrating spheres). Using the above equation, we are able to correct the $\Delta A^*$ signal into a true change in absorption after photoexcitation ($\Delta A$) by measuring the $\Delta R$, $F_R$ and $F_A$ spectra.

We note that in our current experimental setup, it is not possible to measure the change in reflectance (nor the steady-state reflectance spectrum) in our spectroelectrochemical cell – we cannot obtain enough reflected probe light on our detector to do these measurements. We therefore approximate the change in reflectance at a certain applied potential with the change in reflectance at the open circuit potential. We think this is justified, since the band-edge $\Delta A$ amplitude at the open circuit potential is not changed significantly, whereas the sub-bandgap signal (mostly caused by changes in reflection of the sample) is reduced to zero after the correction.

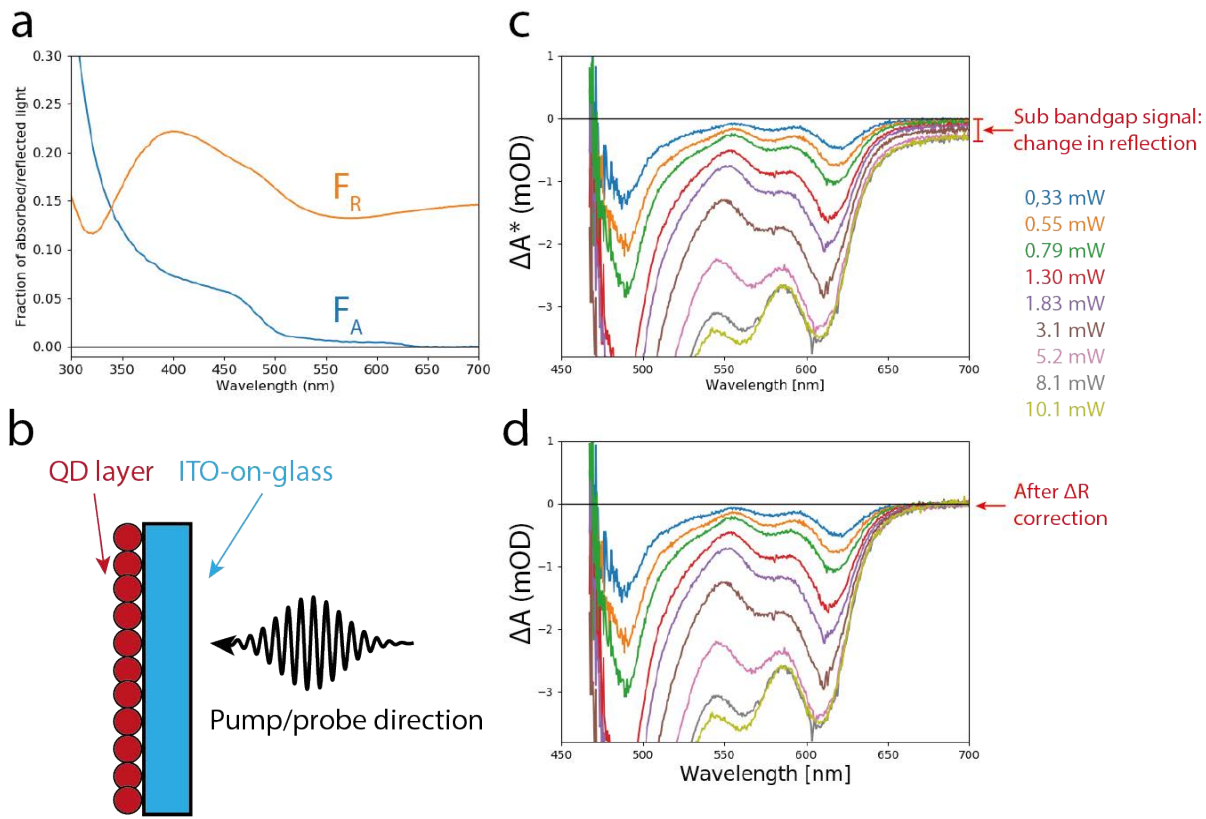

**Figure S9: Reflection correction on a film of QDs. (a)** Steady-state absorption spectrum, plotted as fraction of absorbed light (blue curve), and steady-state reflection spectrum, plotted as fraction of reflected light (yellow curve), as measured inside an integrating sphere. **(b)** Orientation of the QD film w.r.t. the incoming pump and probe beams in *all* transient absorption/reflection experiments. **(c)** Uncorrected transient 'exctinction' spectra of a QD film at the open circuit potential (neutral QD film) at a pump-probe delay time of 5 ps. **(d)** Reflection-corrected transient absorption data, taken the data from **(c)**.

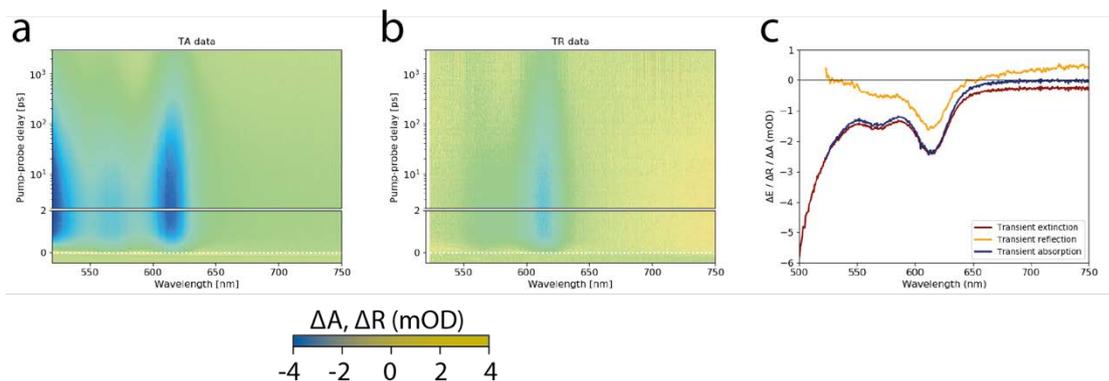

**Figure S10: Reflection correction on a film of QDs – example.** **(a)** As-measured transient-'extinction' data. **(b)** Transient reflection measurement. **(c)** Spectral slices showing the effect of the reflection correction at a pump-probe delay time of 5 ps. The fraction of absorbed and reflected light from Figure S7(a) were used (same QD film). Note that the spectral window we can reach for the correction is limited by the probe spectrum we reflect onto our detector in the transient reflection measurement (yellow curve). Pump-power used for these measurements was 3.1 mW.

# Solution TA data

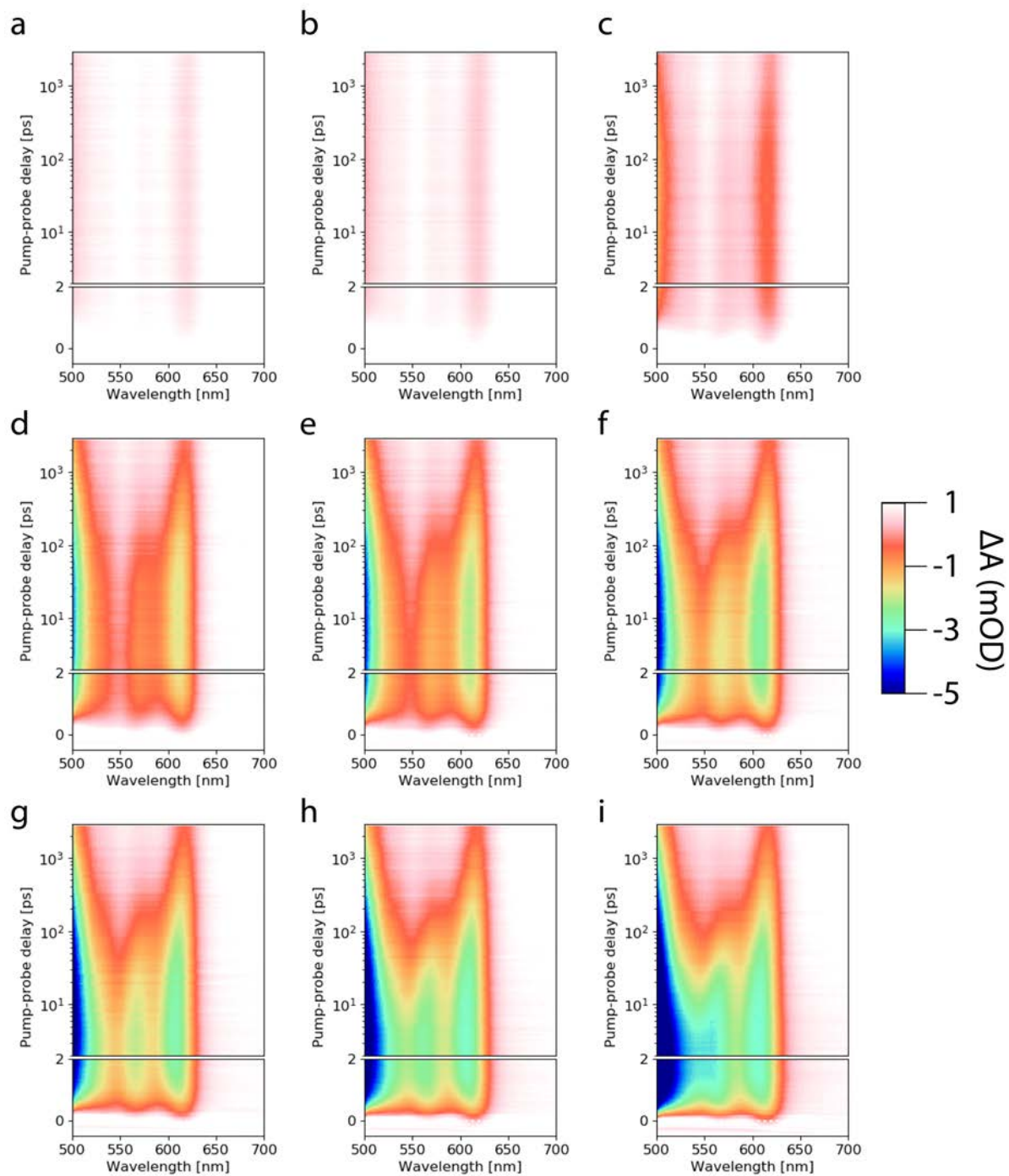

**Figure S11: Fluence dependent TA data from the QDs dispersed in hexane.** (a-g) increasing fluence going from <$N_X$> = 0.34, 0.45, 1.09, 2.66, 3.52, 4.28, 5.03, 7.91, 9.09 respectively. For <$N_X$> determination, see below.

# Absorption cross-section determination in solution

We measure the absorption cross section of the CdSe/CdS/ZnS nanocrystals following Poissonian excitation statistics. The magnitude of the bleach is smaller than directly after photoexcitation, but nonzero, meaning that there still is a finite population of nanocrystals which have excitons in there. The amplitude of the absorption bleach scales with the number of excitons present, and can be estimated via;

$$|\Delta A_{1-3\,ns}| \propto 1 - P_0 = 1 - e^{-\langle N \rangle} \qquad \text{(eq. S34)}$$

Where $|\Delta A_{1-3ns}|$ is the magnitude of the bleach between 1-3 nanoseconds, $P_0$ is the Poisson probability of finding zero excitons and <N> is the average exciton population per nanocrystal. In turn, <N> = σ $J_0$, with σ being the absorption cross section, and $J_0$ the incoming photon fluence. By fitting the data to the above equation, we obtain an absorption cross section at 400 nm of $3.6 \pm 0.2 \cdot 10^{-14}$ cm², which we use to calculate <N> per photon fluence used in the TA experiments.

We also correct the incoming photon fluence $J_0$ for absorption throughout the solution:

$$J_0' = \frac{1 - e^{-\alpha L}}{\alpha L} J_0,$$

with $J_0'$ the average of the photon fluence across the solution length and $\alpha$ the absorption coefficient at the excitation wavelength. The term $\alpha \cdot L$ equals $A \cdot \ln(10)$, with $A$ the optical density at 400 nm. The analysis is shown in Figure S12.

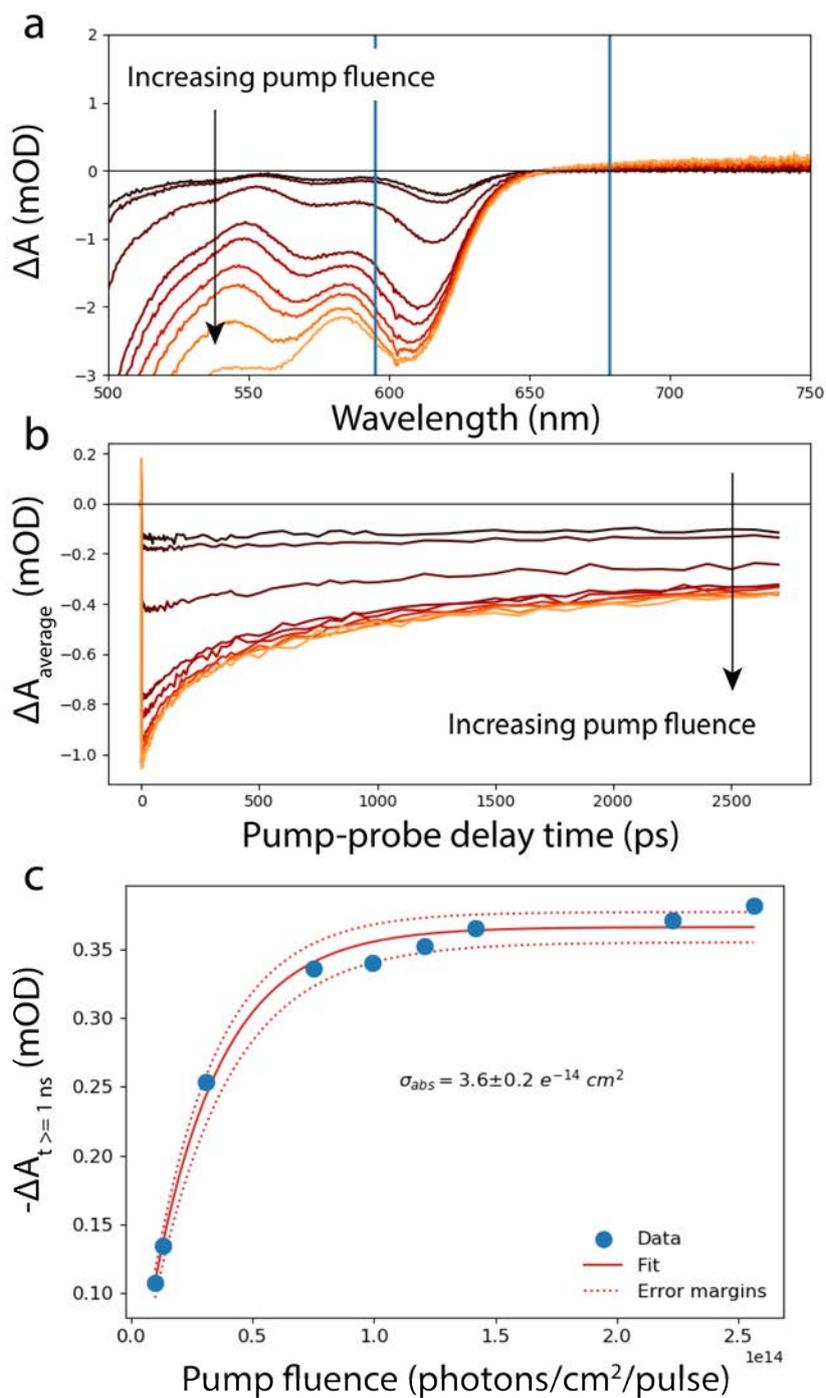

**Figure S12: Absorption cross section at 400 nm determination.** (a) ΔA at 5 ps for increasing pump fluence. The blue vertical lines show the region in which the signal is averaged to obtain the time dependent data in (b). The average bleach after 1 ns is averaged, which is shown in (c) for increasing fluence and fitted with equation S34 to obtain the absorption cross section at the pump wavelength (400 nm).

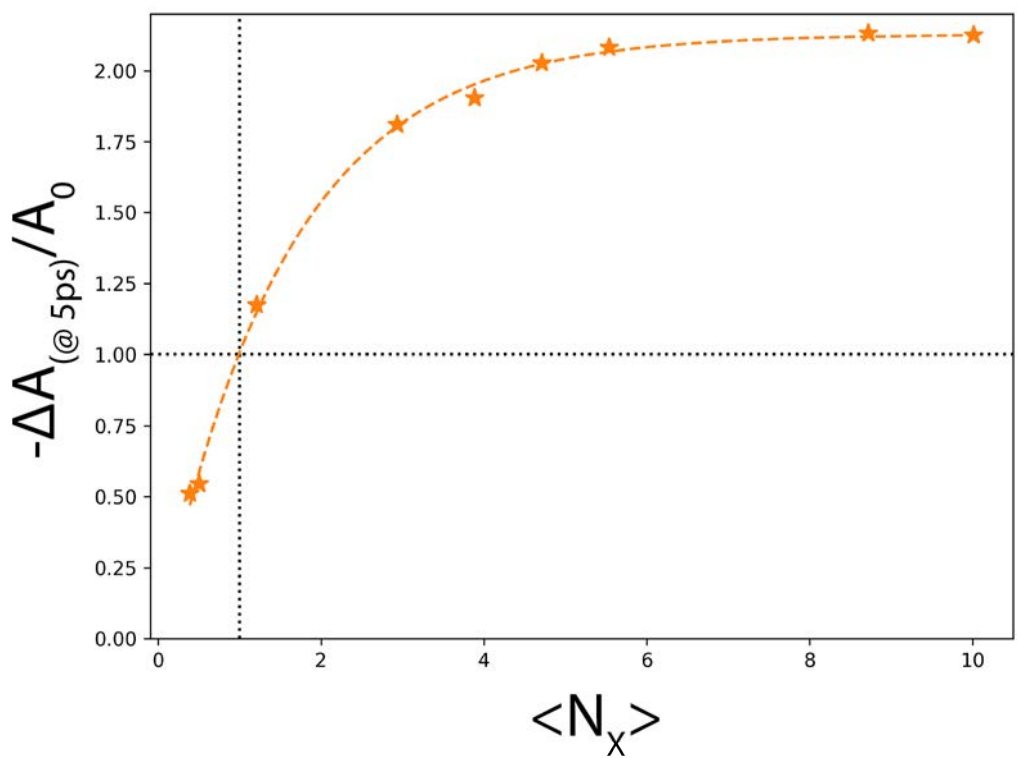

**Figure S13: Fractional bleach as a function of increasing pump fluence.** Data averaged as depicted in Figure S12. Note that the fractional bleach value of 2 means that the absorption spectrum is completely inverted.

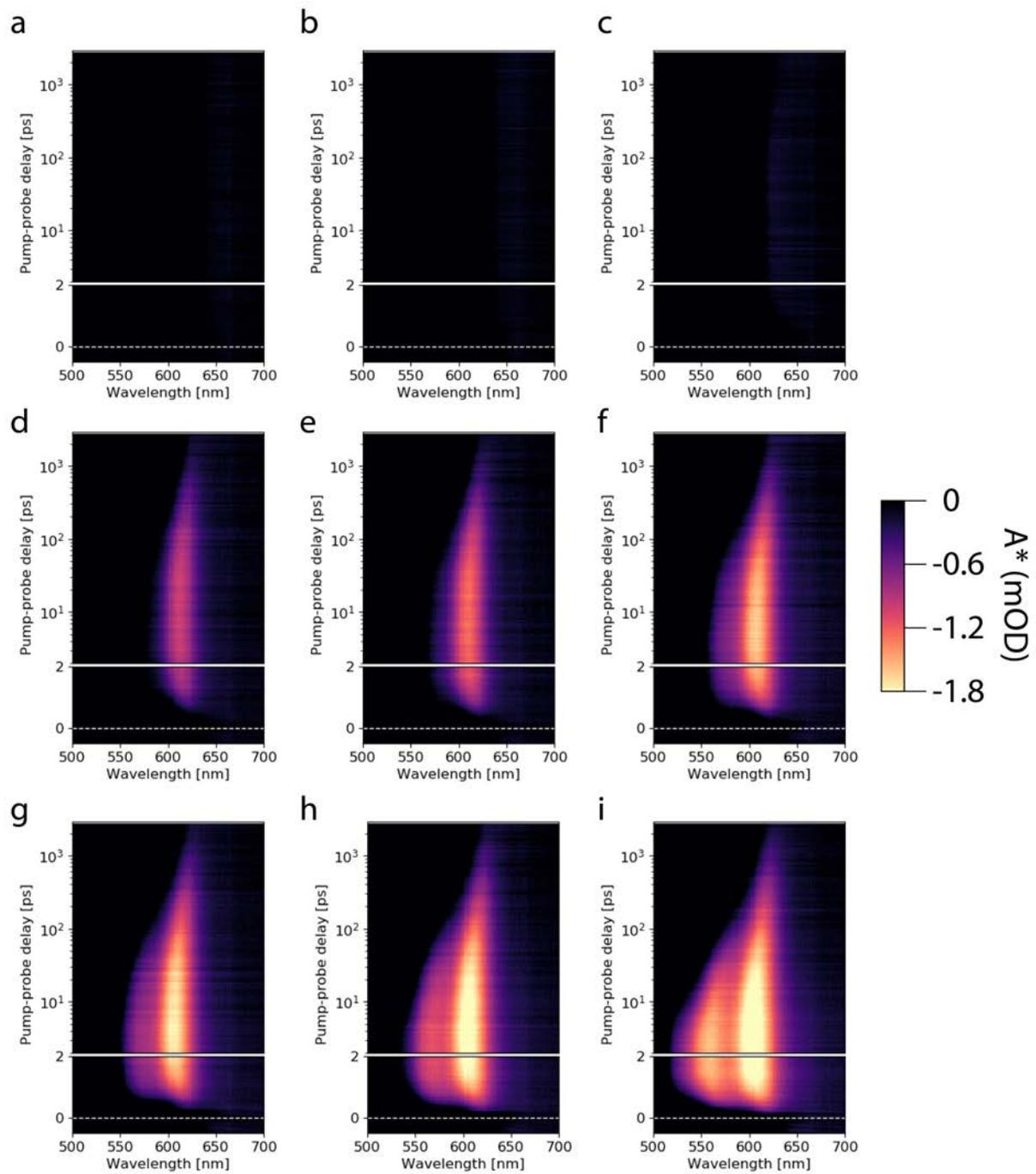

**Figure S14: Fluence dependent gain maps ($A_0 + \Delta A$) from the QDs dispersed in hexane.** (a-g) increasing fluence going from $\langle N_X \rangle =$ 0.34, 0.45, 1.09, 2.66, 3.52, 4.28, 5.03, 7.91, 9.09 respectively. Note that at higher fluences, on the blue side of the 1S transition, also the 2S transition starts to show optical gain.

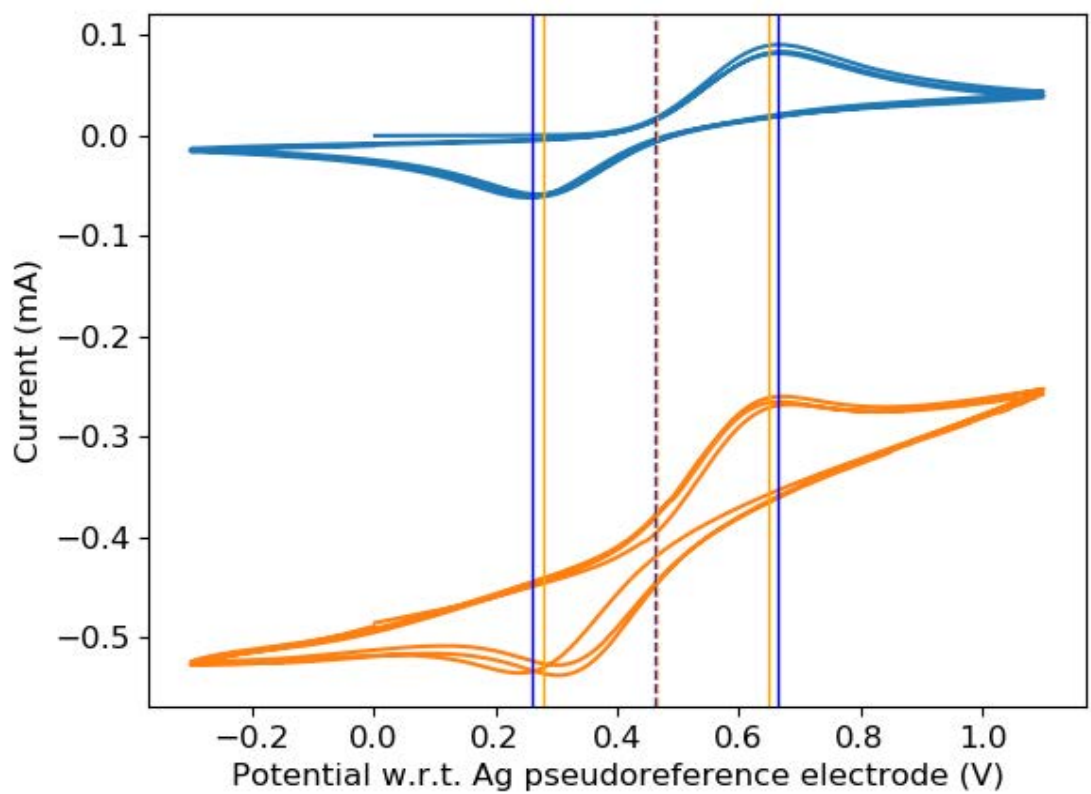

**Figure S15: Cyclic voltammetry of Ferrocene/Ferrocenium before (blue) and after (orange) spectroelectrochemistry + TA to calibrate the Ag pseudoreference electrode.** The orange CV is offset vertically for clarity. All spectroelectrochemical measurements (Figure 2 of the main text) and ultrafast TA measurements (Figure 3 and beyond of the main text) were done in series directly after each other. The blue CV presented here was measured before the all measurements, the orange curve directly after the measurements, after placing the cell back inside an $N_2$ purged glovebox. The vertical lines represent the oxidation and reduction maximum, whereas the dashed line shows the $E_{1/2}$ potential. There is a small shift of 2.5 mV before and after the measurement. The work electrode used in these measurements is a blank ITO-on-glass slide (see methods section of the SI).

## TA data film

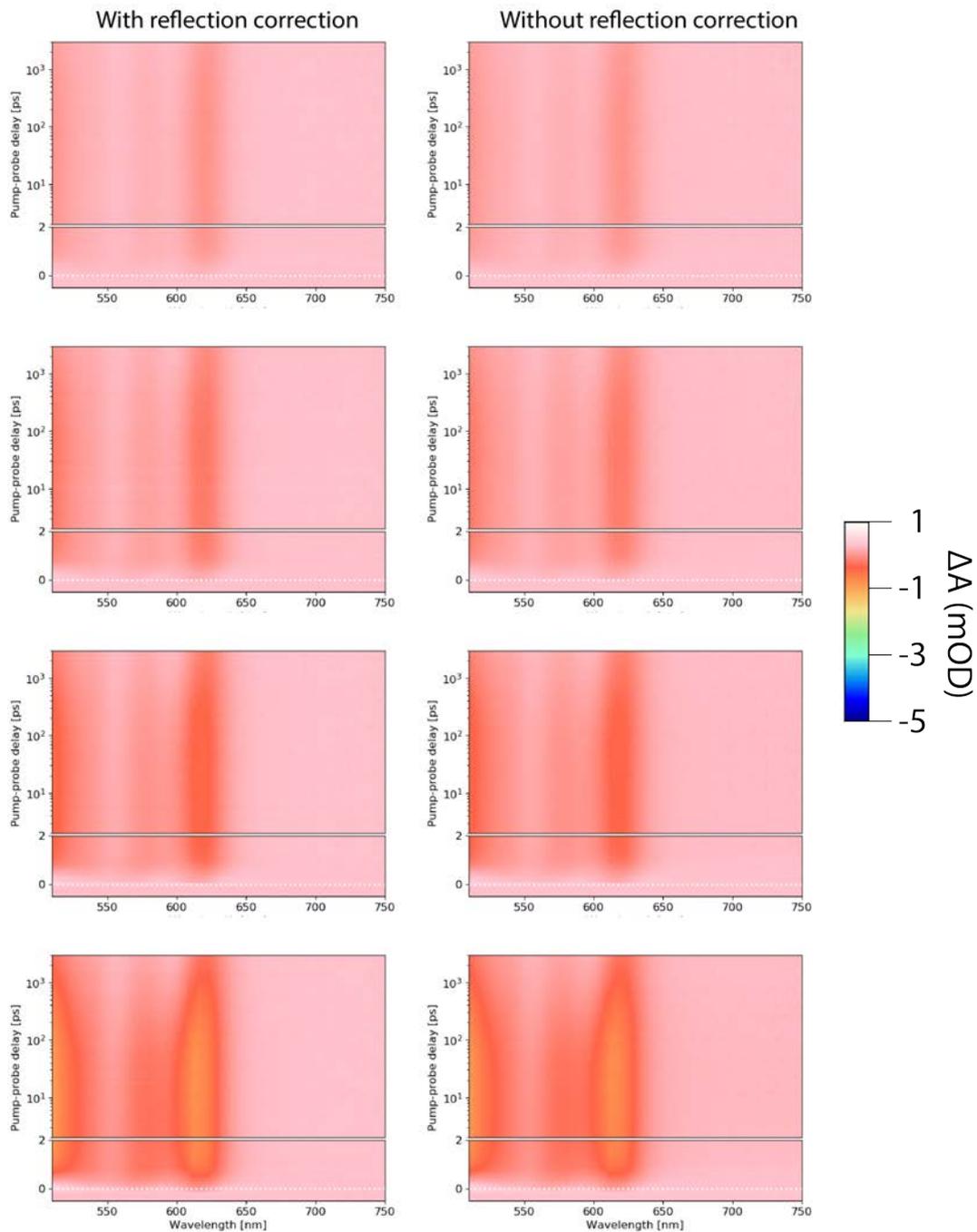

**Figure S16: Reflection uncorrected (left images) and reflection corrected (right images) TA data on film at a doping density of $\langle n_e \rangle = 0$, i.e. at the open circuit potential.** From top to bottom, fluences used produce $\langle N_X \rangle = 0.3, 0.5, 0.8, 1.3$. Note that the effect of the correction becomes more visible at higher fluences.

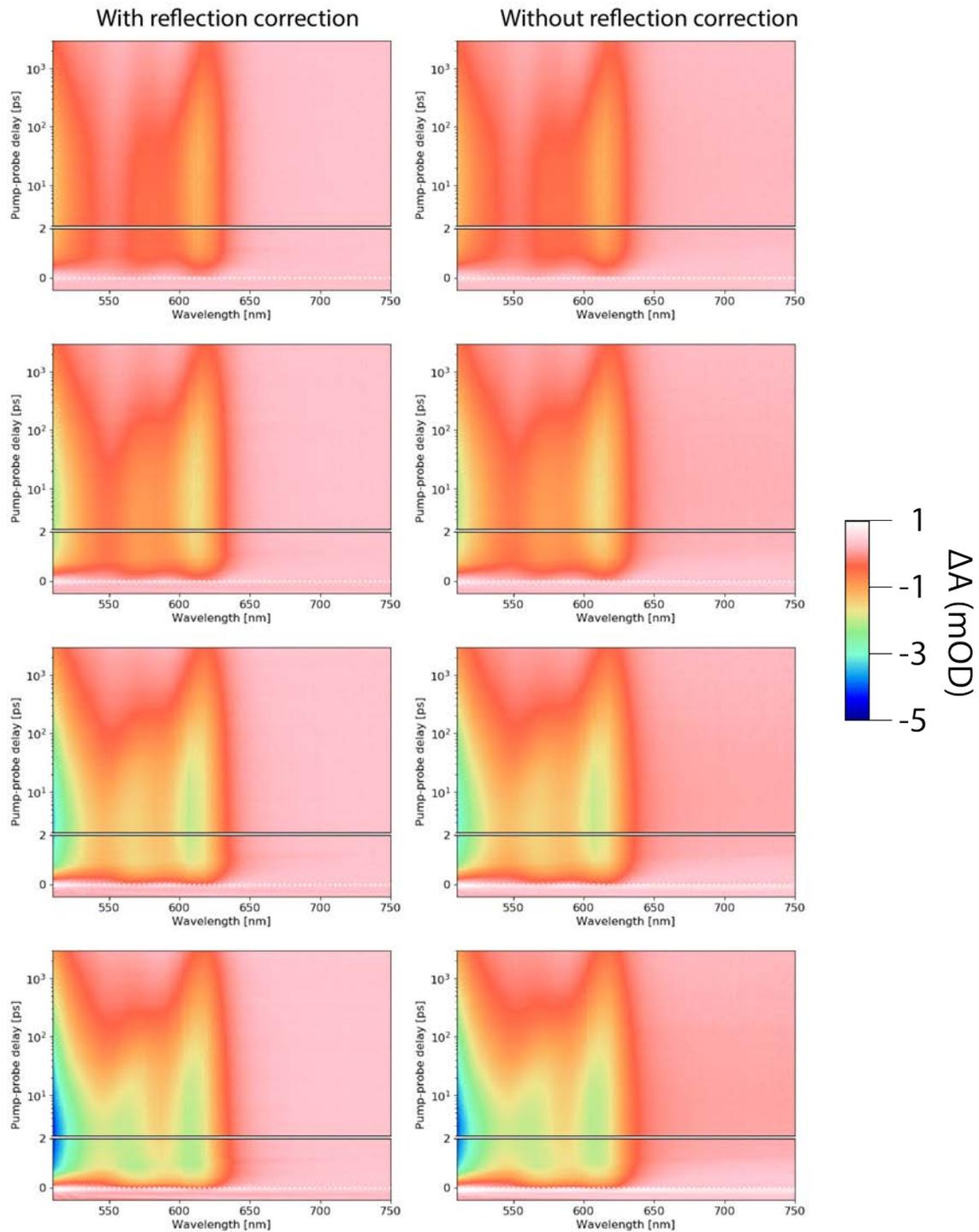

**Figure S17: Reflection uncorrected (left images) and reflection corrected (right images) TA data on film at a doping density of <ne> = 0, i.e. at the open circuit potential.** From top to bottom, fluences used produce $<N_X>$ = 1.9, 3.1, 5.3, 5.9. Note that the effect of the correction becomes more visible at higher fluences.

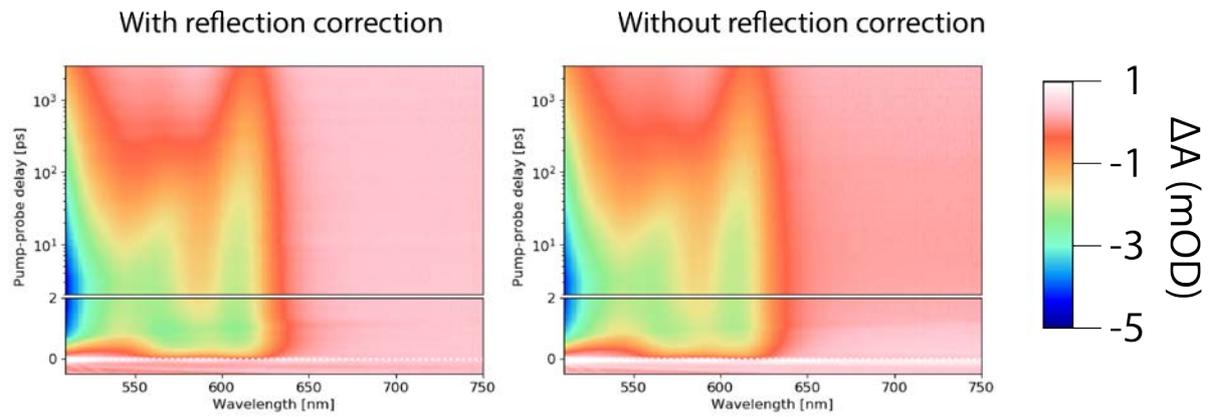

**Figure S18: Reflection uncorrected (left images) and reflection corrected (right images) TA data on film at a doping density of $\langle n_e \rangle = 0$, i.e. at the open circuit potential.** Fluence equals $\langle N_X \rangle = 6.6$. On the non-reflection corrected TA image on the right, a clear shade in the sub-bandgap region (>650 nm) can be observed, which is corrected for in the left TA image. This can be observed like a sub-bandgap bleach in a spectral slice.

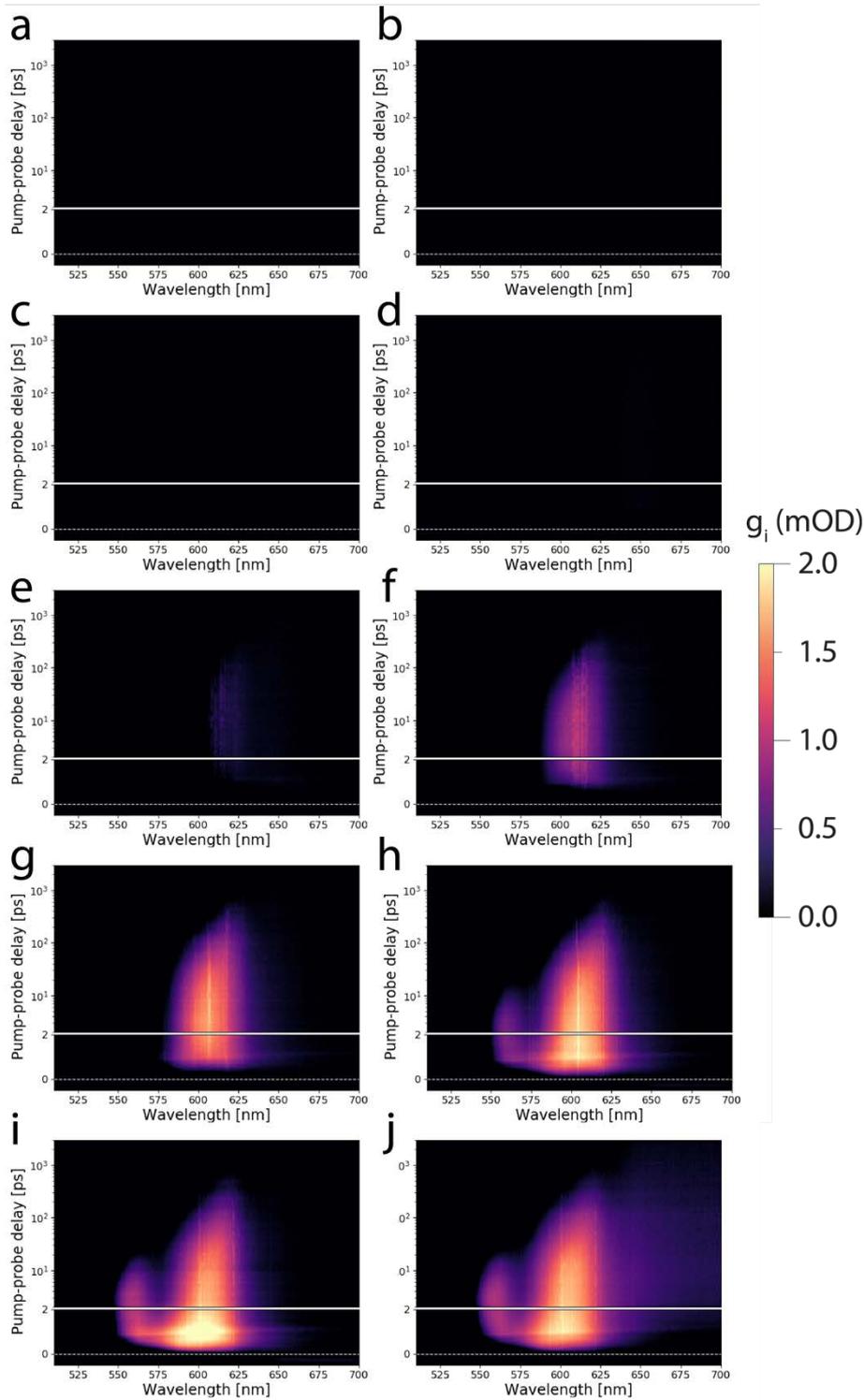

**Figure S19: Gain maps at $\langle n_e \rangle = 0$ for various fluences.** (a-i) fluences are $\langle N_X \rangle$ = 0.3, 0.5, 0.8, 1.3, 1.9, 3.1, 5.3, 5.9, and 6.6 respectively. (j) Same as in (i) but without the background correction. Note the visible sub-bandgap optical gain signal which is effectively removed using the reflection corrected data.

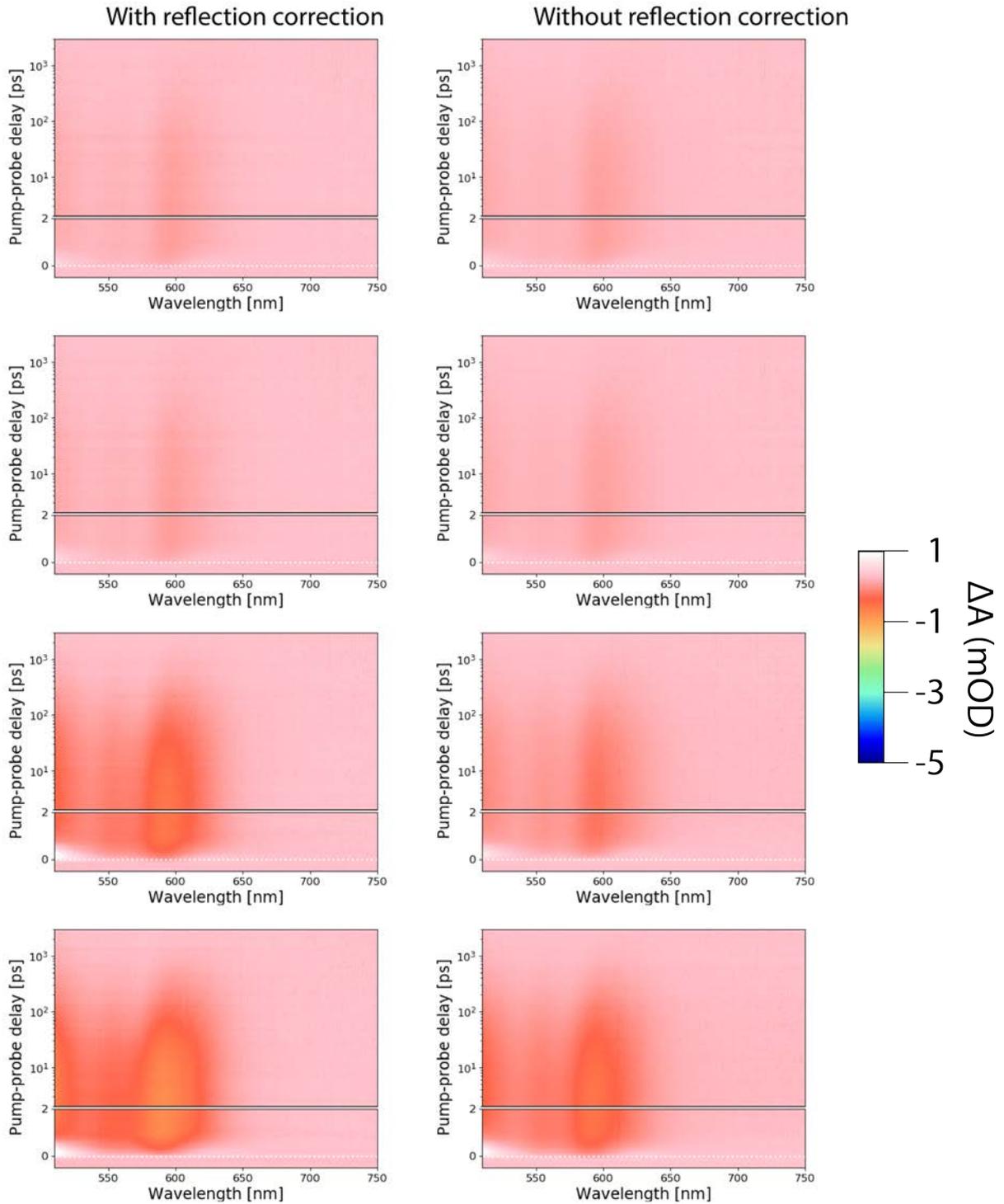

**Figure S20: Reflection uncorrected (left images) and reflection corrected (right images) TA data on film at a doping density of $\langle n_e \rangle = 1.99$.** From top to bottom, fluences used produce $\langle N_X \rangle = 0.3, 0.5, 0.8, 1.3$. Note that the effect of the correction becomes more visible at higher fluences.

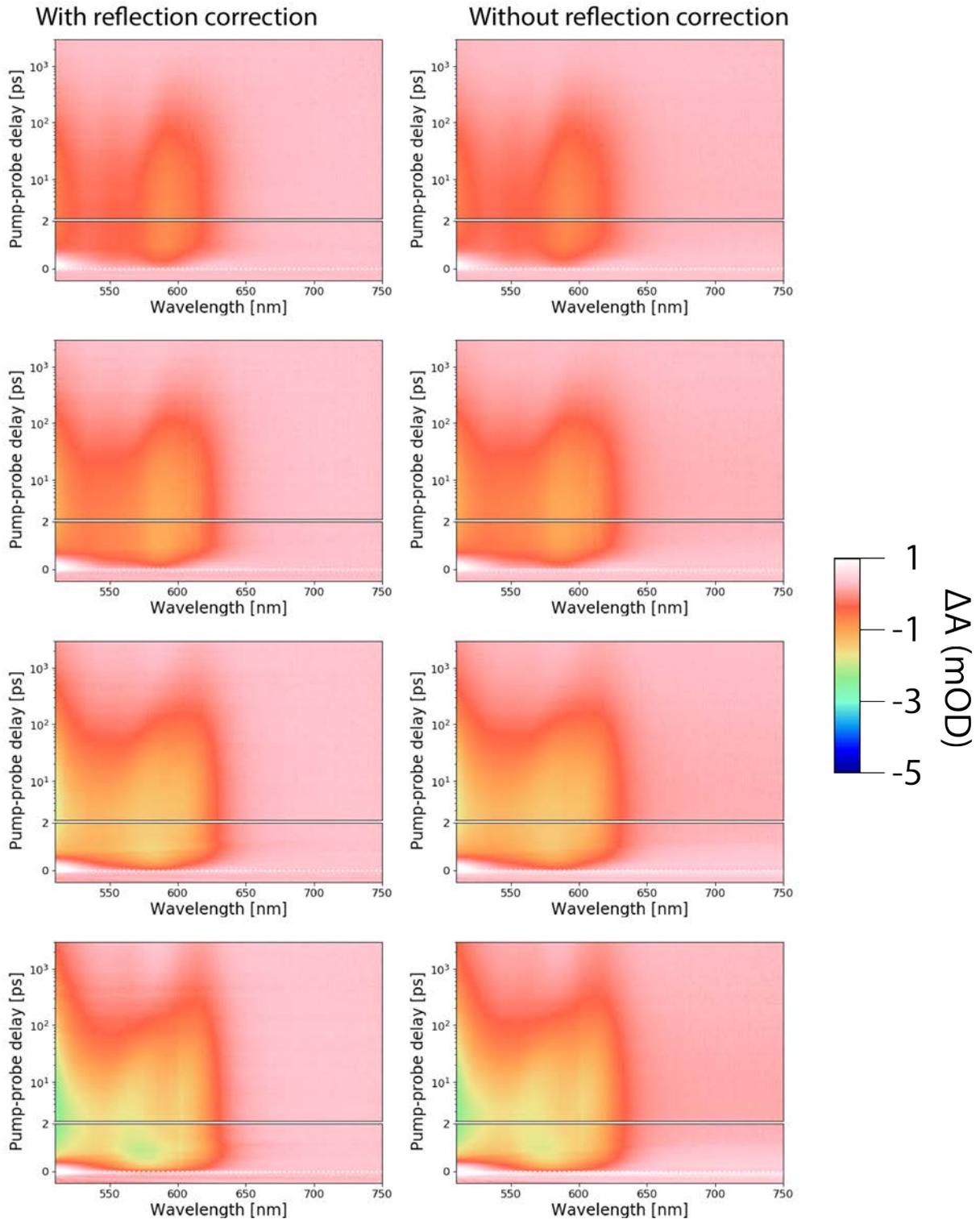

**Figure S21: Reflection uncorrected (left images) and reflection corrected (right images) TA data on film at a doping density of $\langle n_e \rangle$ = 1.99.** From top to bottom, fluences used produce $\langle N_X \rangle$ = 1.9, 3.1, 5.3, 5.9. Note that the effect of the correction becomes more visible at higher fluences.

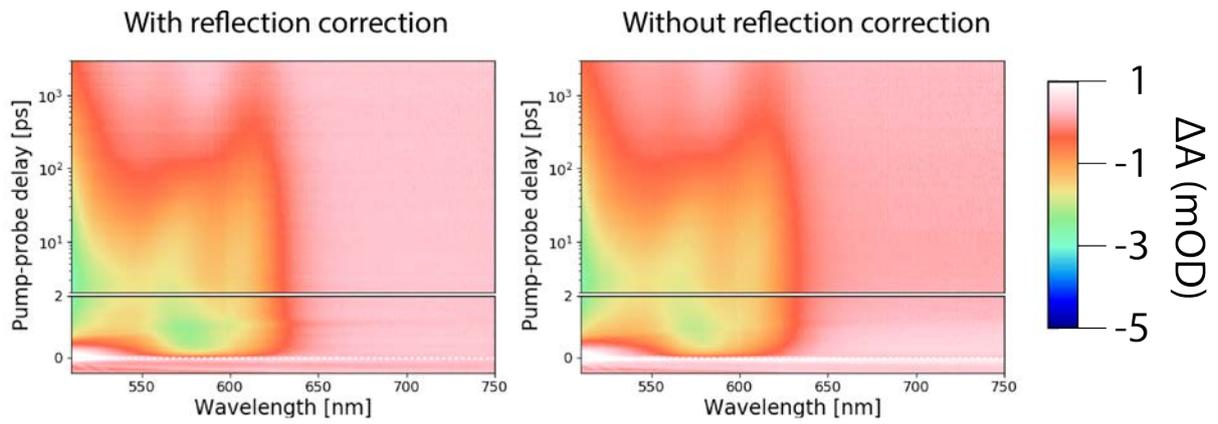

**Figure S22: Reflection uncorrected (left images) and reflection corrected (right images) TA data on film at a doping density of <n_e> = 0.** Fluence equals $<N_X>$ = 6.6. On the non-reflection corrected TA image on the right, a clear shade in the sub-bandgap region (>650 nm) can be observed, which is corrected for in the left TA image. This can be observed like a sub-bandgap bleach in a spectral slice.

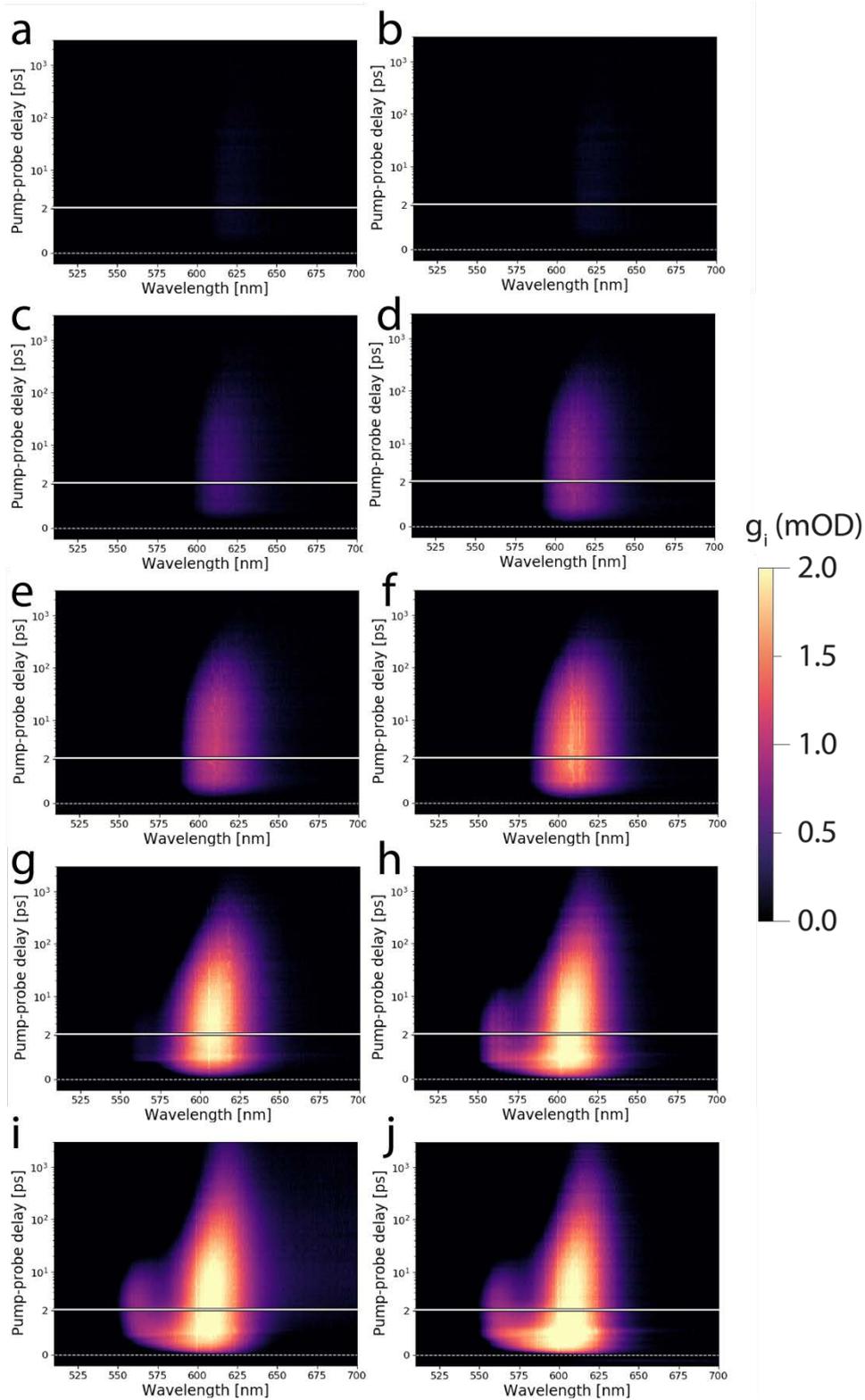

**Figure S23: Gain maps at $\langle n_e \rangle = 0$ for various fluences.** (a-i) fluences are $\langle N_X \rangle$ = 0.3, 0.5, 0.8, 1.3, 1.9, 3.1, 5.3, 5.9, and 6.6 respectively. (j) Same as in (i) but without the background correction.

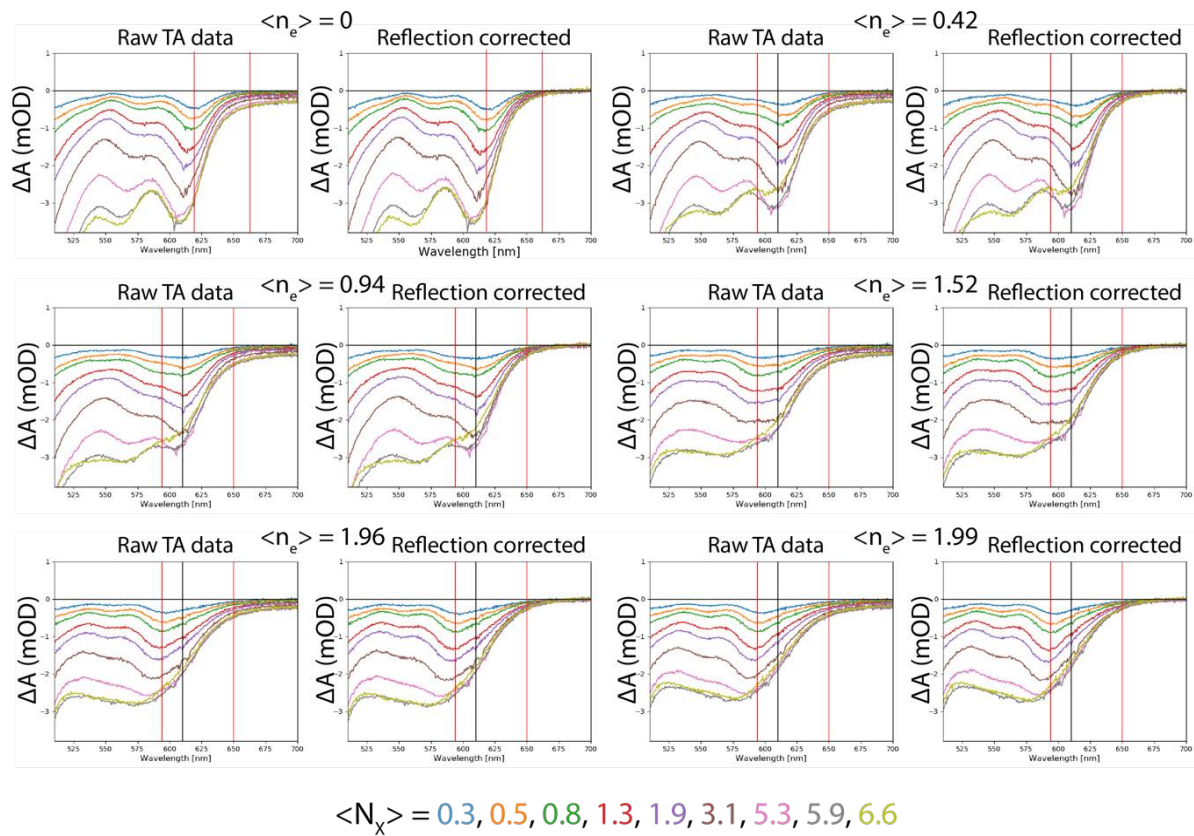

**Figure S24: A spectra at a pump-probe delay time of 5 ps, at varying fluences <N$_X$> and doping densities <n$_e$>.** Left panels show the raw TA data, right panels the TA data that is corrected for reflection. Note how the sub-bandgap apparent bleach is effectively removed.

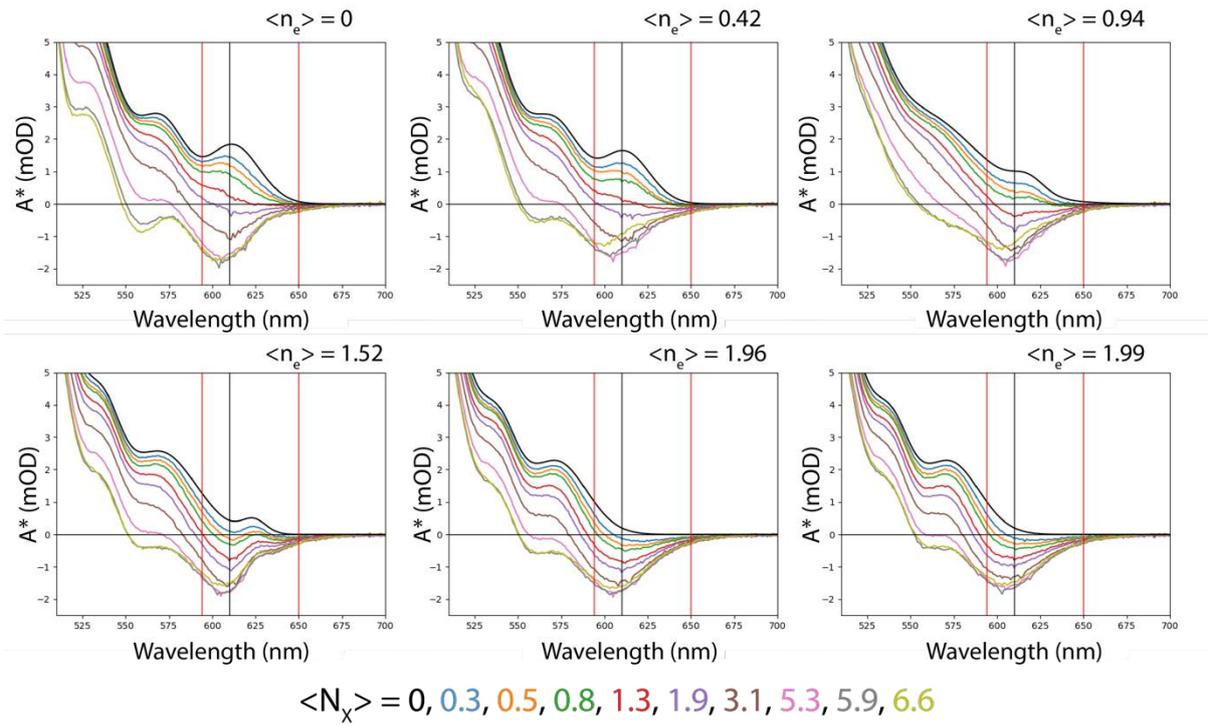

**Figure S25: Excited state absorption spectra A\* ($A_0 + \Delta A_{TA} + \Delta A_{SEC}$) spectra at a pump-probe delay time of 5 ps, at varying fluences $\langle N_X \rangle$ and doping densities $\langle n_e \rangle$.**

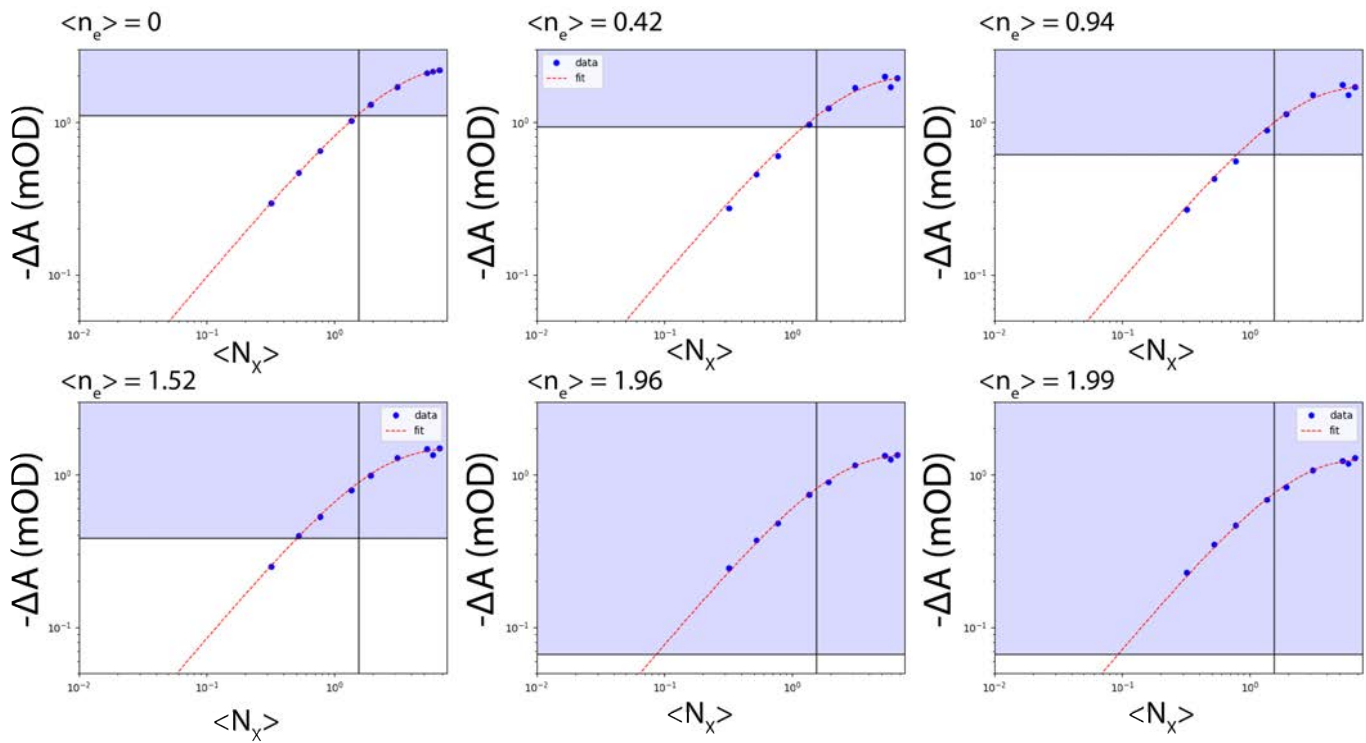

**Figure S26: Gain threshold determination for various doping densities $\langle n_e \rangle$ and fluences $\langle N_x \rangle$.** The blue datapoints show the averaged bleach amplitude over the 1S transition, the red dashed line a fit in order to determine the crossing point. The horizontal black line is the average absorption over the 1S transition, the vertical black line is the theoretical gain threshold $\langle N_{gain} \rangle$ = 1.54 for undoped CdSe QDs. The gain thresholds for increasing doping densities are $\langle N_{gain} \rangle$ = 1.51, 1.25, 0.80, 0.51, 0.09, and 0.09 excitons per QD.

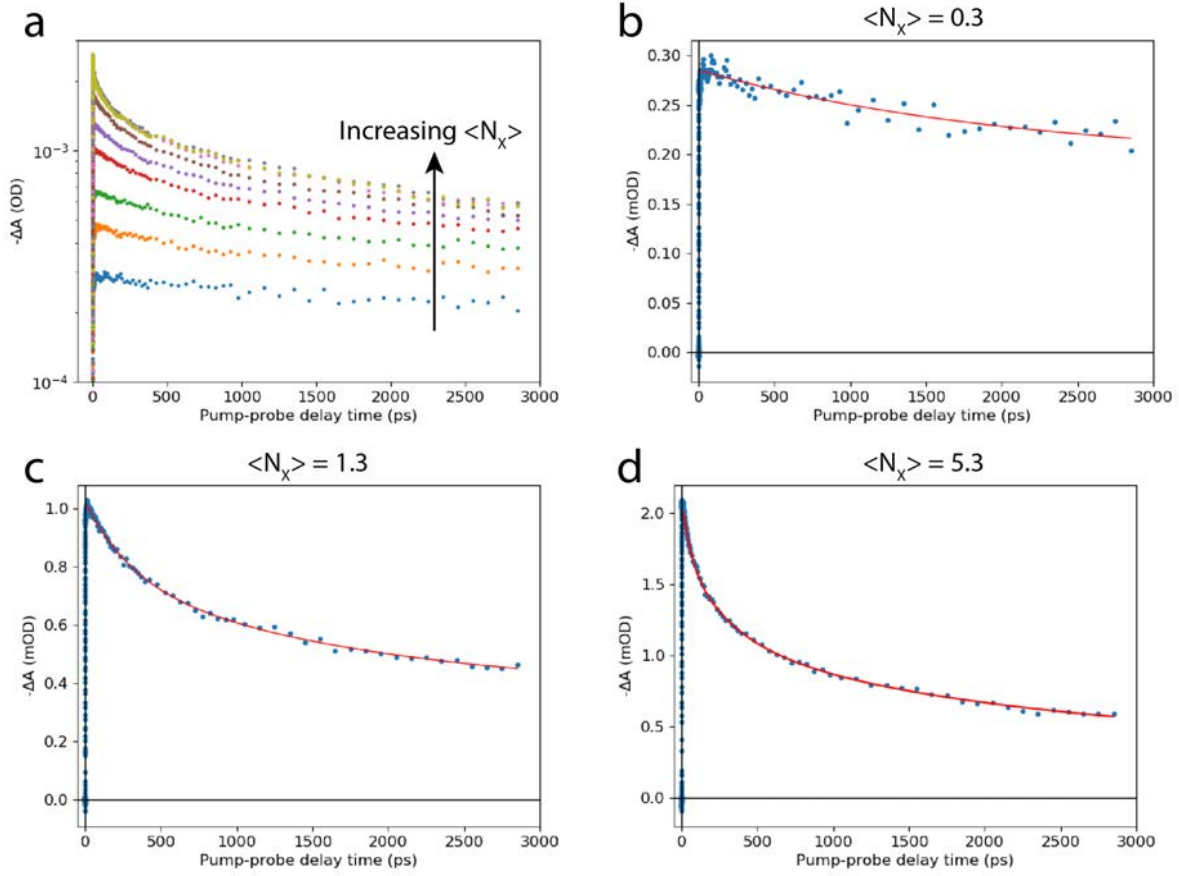

**Figure S27: Determination of the biexciton and triexciton lifetime from TA measurements on the neutral QD film.** (a) Spectrally integrated band-edge bleach, $-\Delta A_{TA}$, as a function of pump-probe delay time for fluences $\langle N_X \rangle$ = 0.3, 0.5, 0.8, 1.3, 1.9, 3.1, 5.3, 5.9, and 6.6. (b) Lowest fluence data, fitted with a single exponential decay. The decay rate was kept constant for fitting the higher fluence data. (c) Data for a fluence of $\langle N_X \rangle$ = 1.3, fitted with a double exponential, yielding a biexciton lifetime of 310 ps. (d) Data for a fluence of $\langle N_X \rangle$ = 5.3, fitted with a triple exponential, yielding a triexciton lifetime of 61 ps.

The lifetime of the multiexcitonic species (biexciton and triexcitons) were obtained by fitting the data with an increasing number of exponentials. The lowest fluence data was fitted with a single exponential decay. The decay rate $r_i$ was fixed for higher order exponential functions, and the amplitude $A_i$ was allowed to vary;

$$-\Delta A(t) = BG + \sum_{i=1}^{3} A_i e^{-r_i t} \qquad \text{(eq. S35)}$$

We used a constant background (*BG*), as the signal does not decay fully in our pump-probe delay time window. We fitted up to three exponential terms for increasing fluences and interpret each increasing term *i* as originating from a higher order excitonic species (i.e. *i* = 1: exciton decay → *i* = 2: biexciton decay → *i* = 3: triexciton decay).

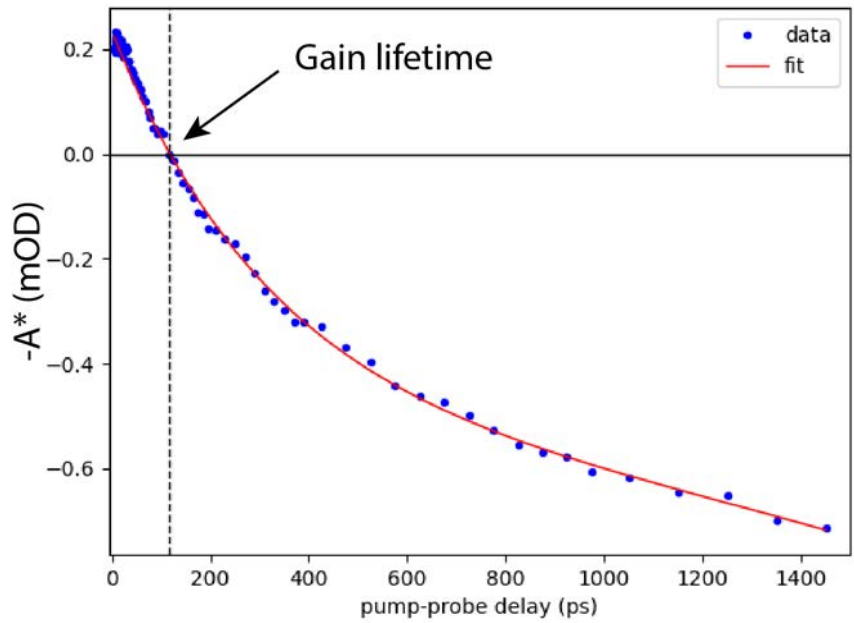

**Figure S28: Determination of the gain lifetime.** (a) Spectrally integrated excited state absorption (from e.g. Figure S23) -A* as a function of pump-probe delay time. We fit with a (sum of) decaying exponential(s) to determine how long the absorption stays negative, which we define as the optical gain lifetime. The example above is for the neutral QD film for a fluence of $\langle N_X \rangle$ = 1.9 excitons/QD. The obtained optical gain lifetimes for different fluences and doping densities are presented in Figure 5(c) of the main text.

# Dataset on second batch of QDs with lower PLQY

We measured a complete dataset on QDs with a lower PLQY. The gain threshold and gain threshold reduction are lower than for the batch of QDs used throughout the main text. The hypothesis we propose is that due to the lower PLQY there is more hole-trapping present in these QDs and in the QD film. This increases the gain threshold for the QDs in solution and lowers the threshold reduction in the doped QD solid. The quality, i.e. the PLQY of the starting QDs, is of vital importance for quantitative modelling of the carrier dynamics, but also for the final creation of low-threshold amplifiers and lasers.

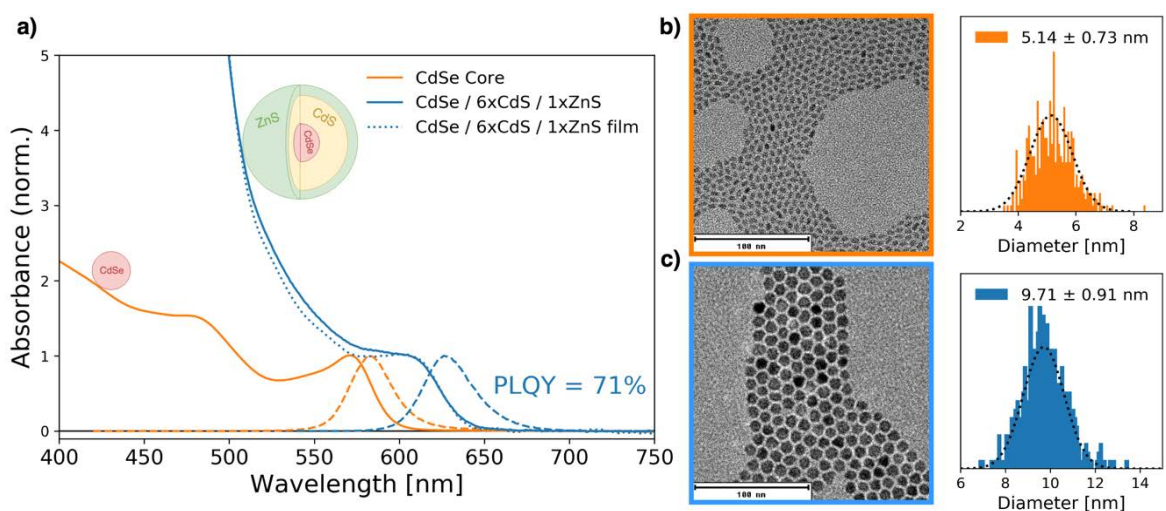

**Figure S29: Optical and structural characterization of the second batch of QDs.** (**a**) Absorption and PL of the core CdSe QDs (orange) and core-shell-shell (blue) QDs. The quantum yield of the final cores is measured to be 71% w.r.t. a reference dye. (**b**) Representative TEM image of the cores, with a histogram of the measured sizes as analyzed with standard ImageJ routines. (**c**) Same as in (**b**), but for the core-shell-shell QDs. Note that we grew only one ZnS layer and that the PLQY is 10% lower than for the QDs presented throughout the main text.

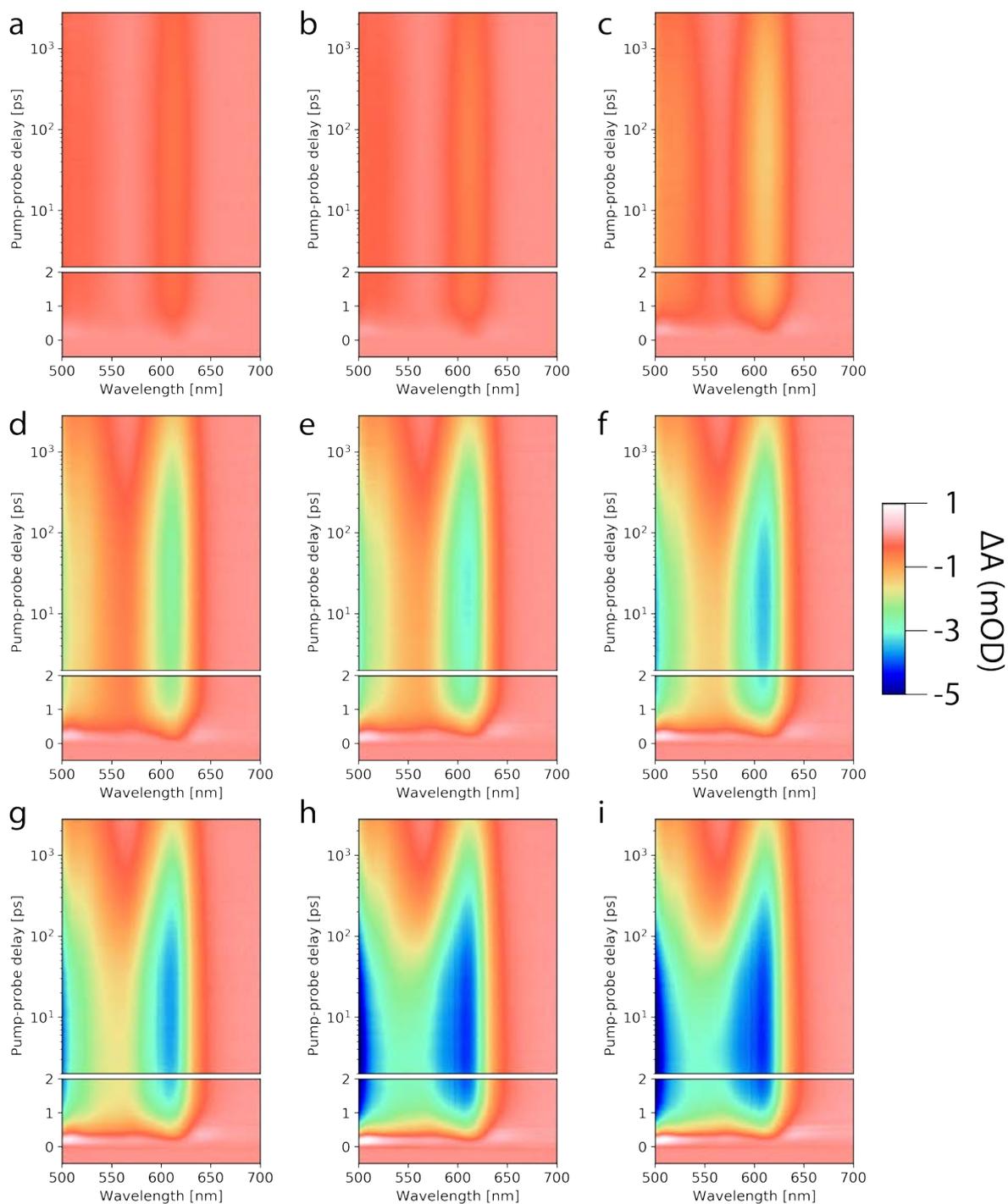

**Figure S30: Fluence dependent TA data from the second batch of QDs dispersed in hexane. (a-g)** increasing fluence going from $<N_X> = $ 0.27, 0.33, 0.73, 1.79, 2.36, 2.92, 3.4, 5.22 and 6.01 respectively.

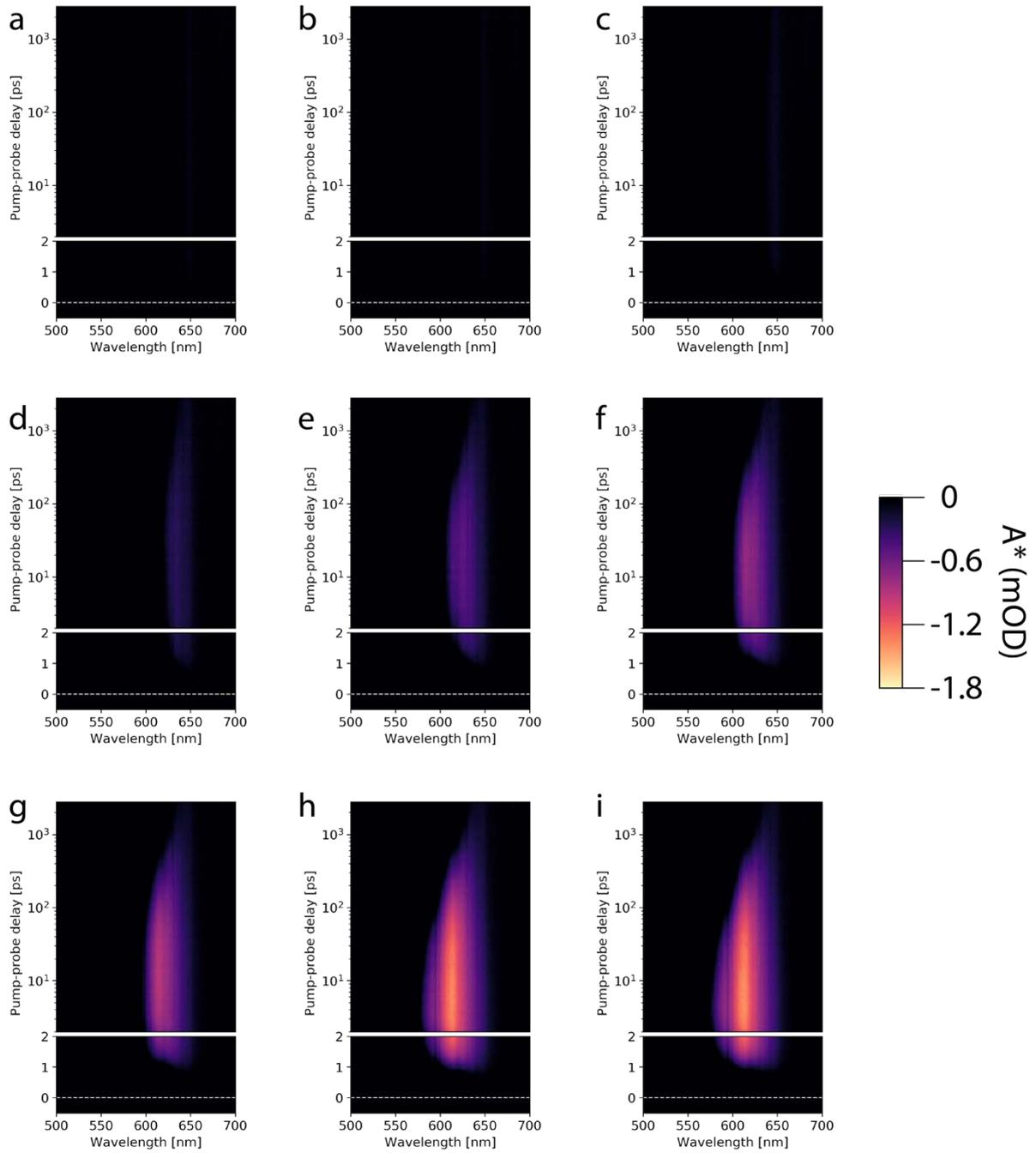

**Figure S31: Fluence dependent gain maps ($A_0 + \Delta A$) from the second batch of QDs dispersed in hexane. (a-g)** increasing fluence going from $\langle N_X \rangle$ = 0.27, 0.33, 0.73, 1.79, 2.36, 2.92, 3.4, 5.22 and 6.01 respectively.

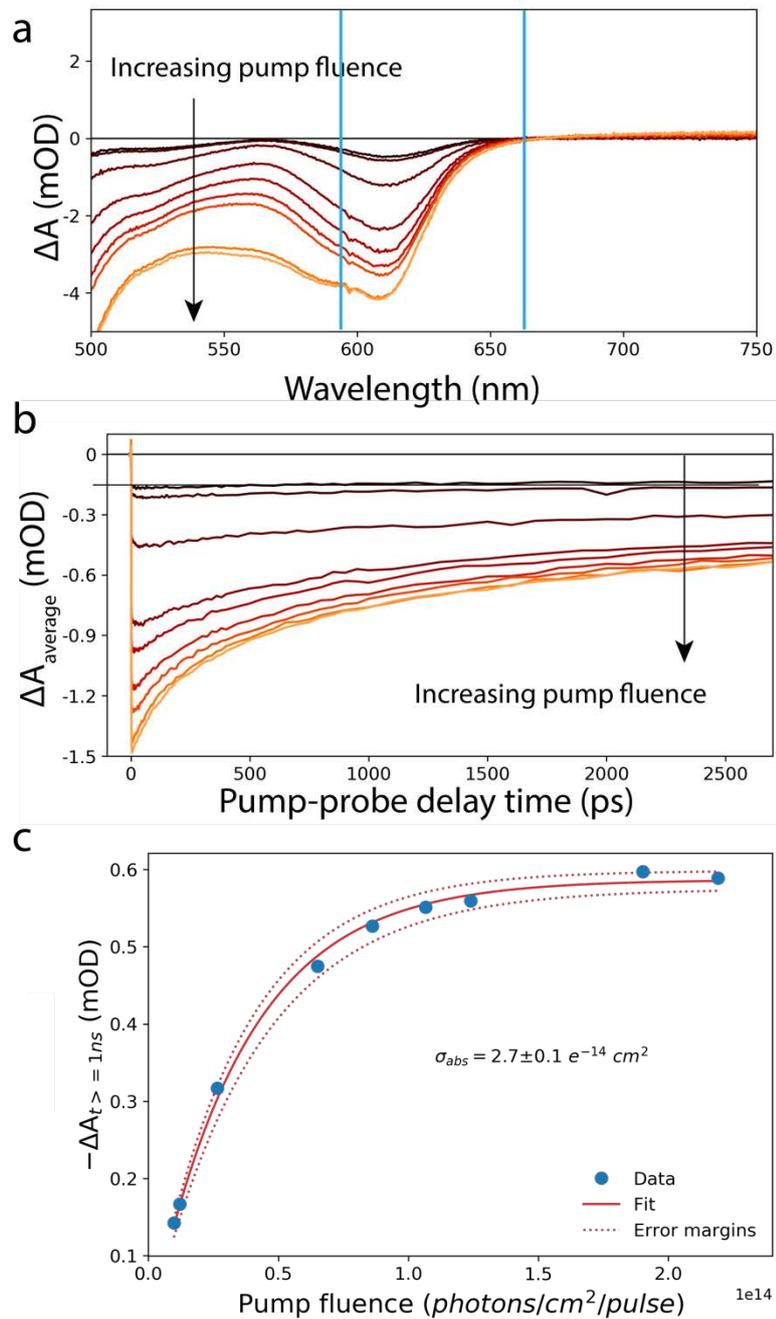

**Figure S32: Absorption cross section at 400 nm determination for the second batch of QDs.** (a) ΔA at 5 ps for increasing pump fluence. The blue vertical lines show the region in which the signal is averaged to obtain the time dependent data in (b). The average bleach after 1 ns is averaged, which is shown in (c) for increasing fluence and fitted with equation S34 to obtain the absorption cross section at the pump wavelength (400 nm).

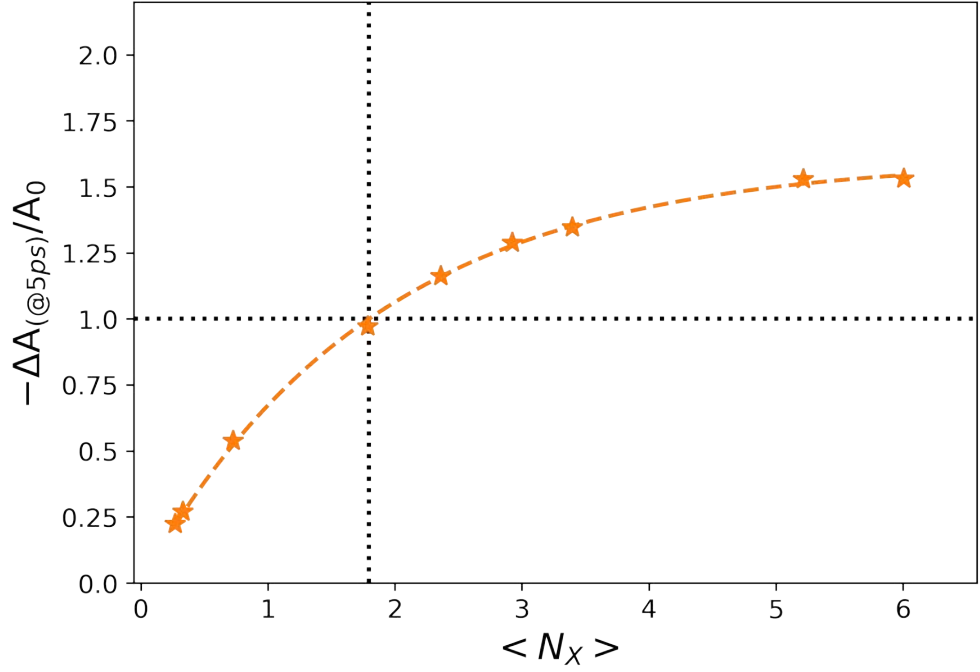

**Figure S33: Fractional bleach as a function of increasing pump fluence for the second batch of QDs.** Data averaged as depicted in Figure S32. Note that the gain threshold is $<N_{gain}> = 1.8$ and we do not fully invert the absorption spectrum. We hypothesize that this is due to a larger amount of hole trapping in these lower-PLQY-QDs.

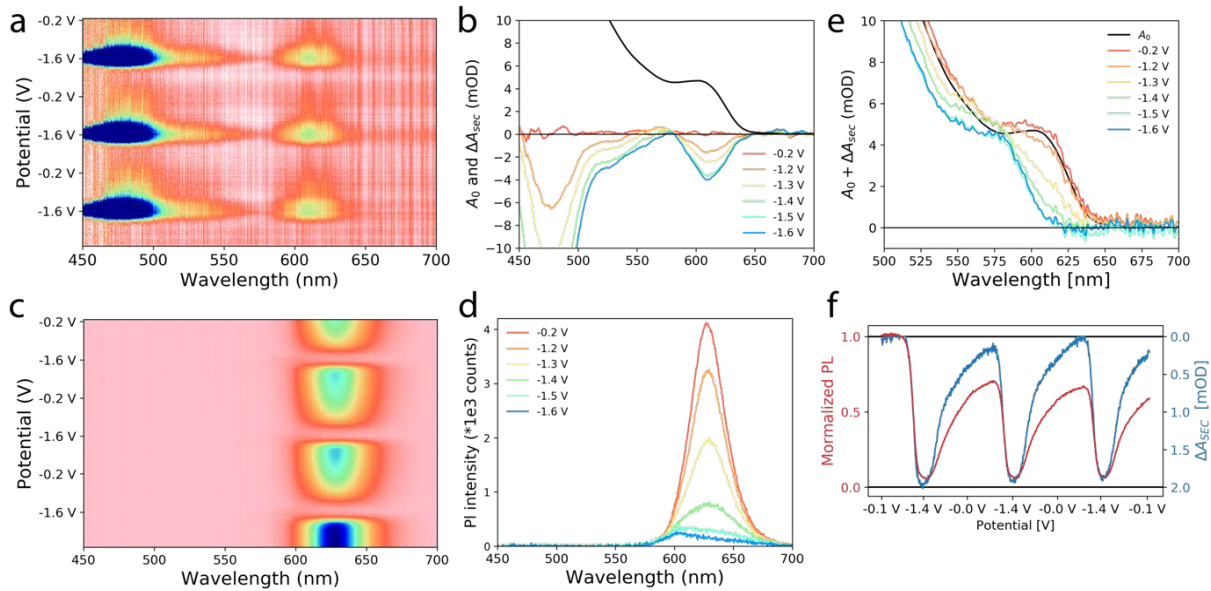

**Figure S34: Spectroelectrochemistry on a film of CdSe/CdS/ZnS from the second batch of QDs.** The potential during all SEC measurements was swept three times between the open circuit potential and -1.6V to check for sample stability. **(a)** Absorption SEC on the QD film. Injection of electrons into the conduction band of the QDs is observed by the bleaching of the band-edge and CdS shell transitions. **(b)** $\Delta A_{SEC}$ spectra at different potentials. Charge injection starts around -1.1V. Note that the band-edge bleach amplitude at the most negative potentials equals the amplitude of the absorption spectrum. **(c)** PL SEC on the QD film. As electrons are injected into the conduction band of the QDs, the PL quenches due to the opening of an Auger recombination channel. **(d)** PL spectra at different potentials. The PL amplitude decreases due to enhanced Auger recombination. At the most negative potential, an additional blue shifted ePL peak appears (negative tetron luminescence). **(e)** Total absorption of the QD film, i.e. $\Delta A_{SEC}+A_0$. The band-edge transition becomes transparent when we inject on average two electrons per QD. **(f)** Normalized PL and amplitude of the band-edge bleach as a function of applied potential. The drop in PL coincides with the injection of charges into the conduction band of the QDs, indicating that we have a relatively trap-free QD film. The amount of electrochemically injected electrons oscillates between zero and two.

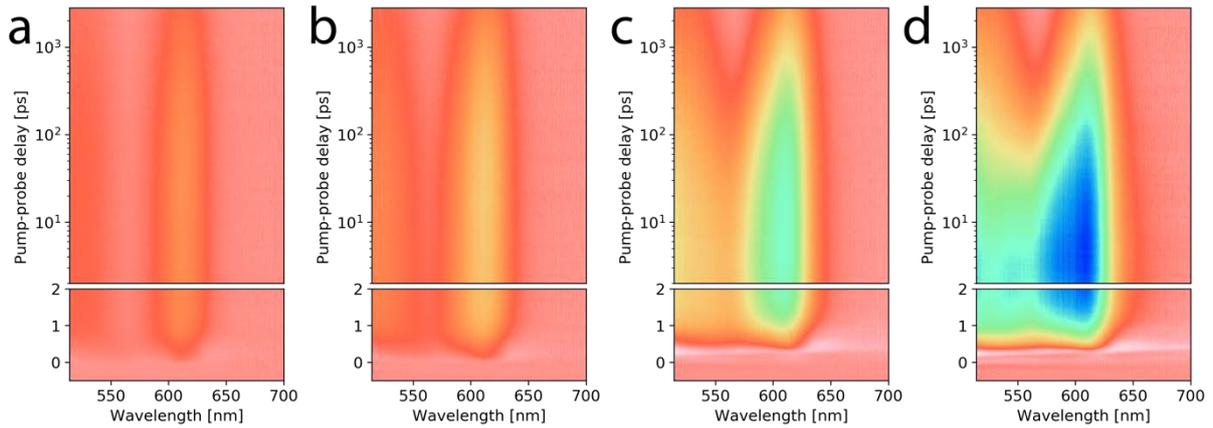

**Figure S35: TA images at -0.2V ($<n_{1S(e)}> = 0$) for increasing fluences.** Going from (a) to (d), we excite $<N_X> = 0.2, 0.5, 1.6$ and $3.2$.

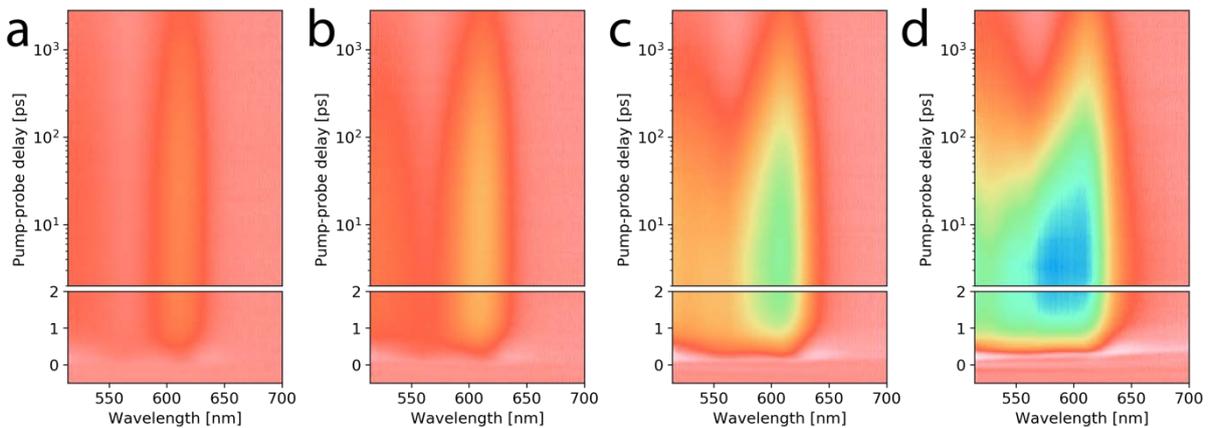

**Figure S36: TA images at -1.2V ($<n_{1S(e)}> = 0.8$) for increasing fluences.** Going from (a) to (d), we excite $<N_X> = 0.2, 0.5, 1.6$ and $3.2$.

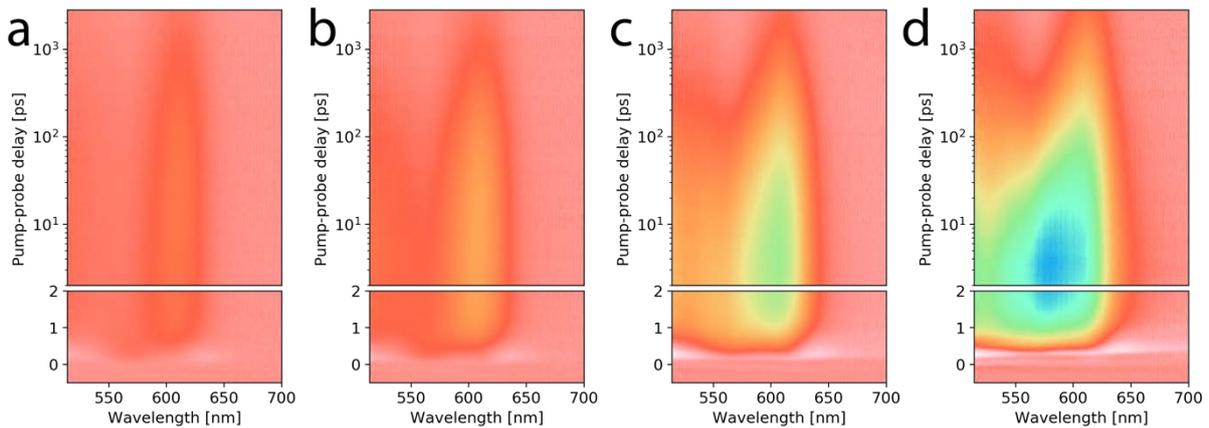

**Figure S37: TA images at -1.3V ($<n_{1S(e)}> = 1.5$) for increasing fluences.** Going from (a) to (d), we excite $<N_X> = 0.2, 0.5, 1.6$ and $3.2$.

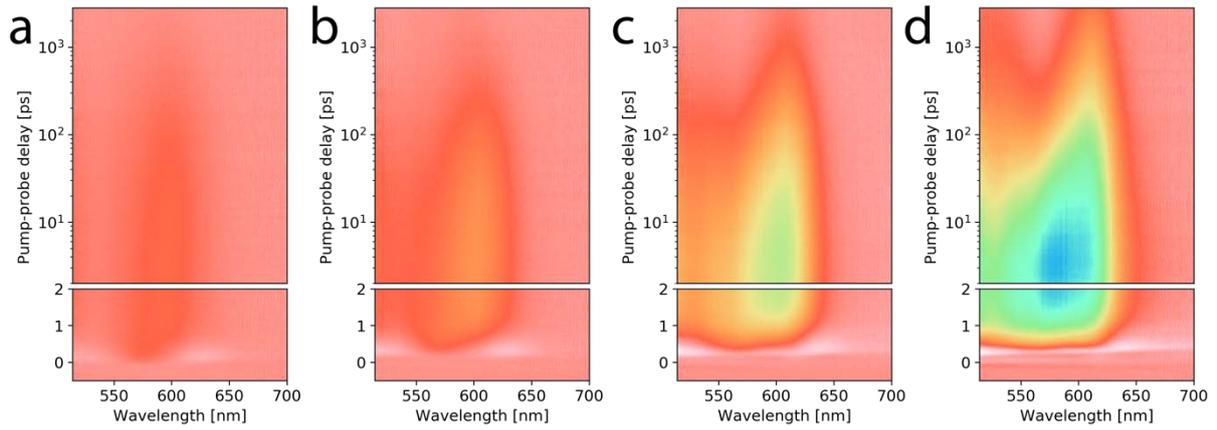

**Figure S38: TA images at -1.4V ($<n_{1S(e)}> = 1.9$) for increasing fluences.** Going from (a) to (d), we excite $<N_X> = 0.2, 0.5, 1.6$ and $3.2$.

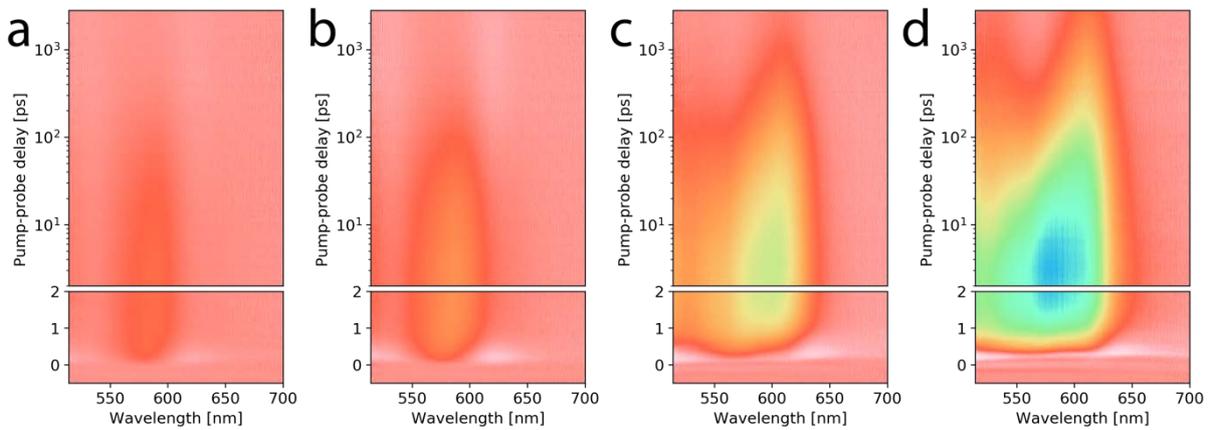

**Figure S39: TA images at -1.5V ($<n_{1S(e)}> = 2.0$) for increasing fluences.** Going from (a) to (d), we excite $<N_X> = 0.2, 0.5, 1.6$ and $3.2$.

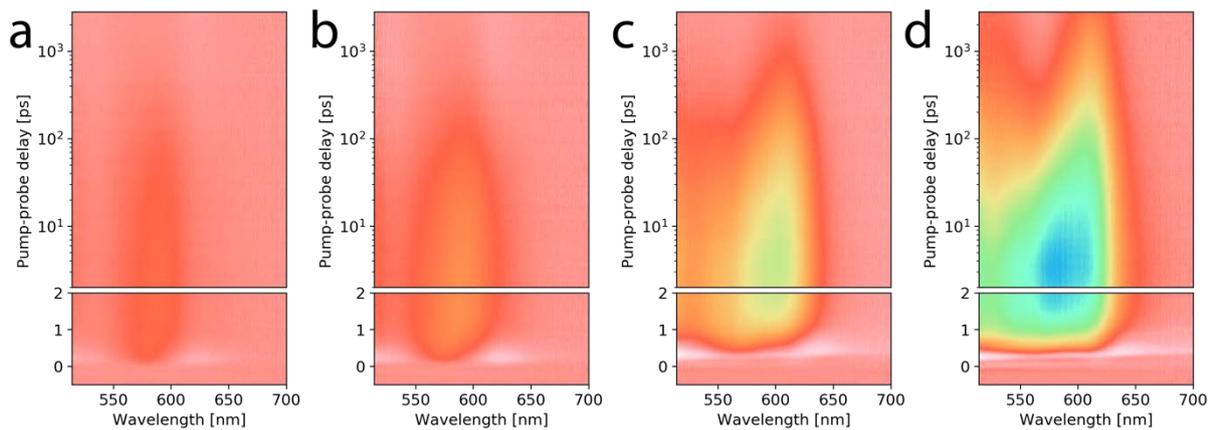

**Figure S40: TA images at -1.6V ($<n_{1S(e)}> = 2.0$) for increasing fluences.** Going from (a) to (d), we excite $<N_X> = 0.2, 0.5, 1.6$ and $3.2$.

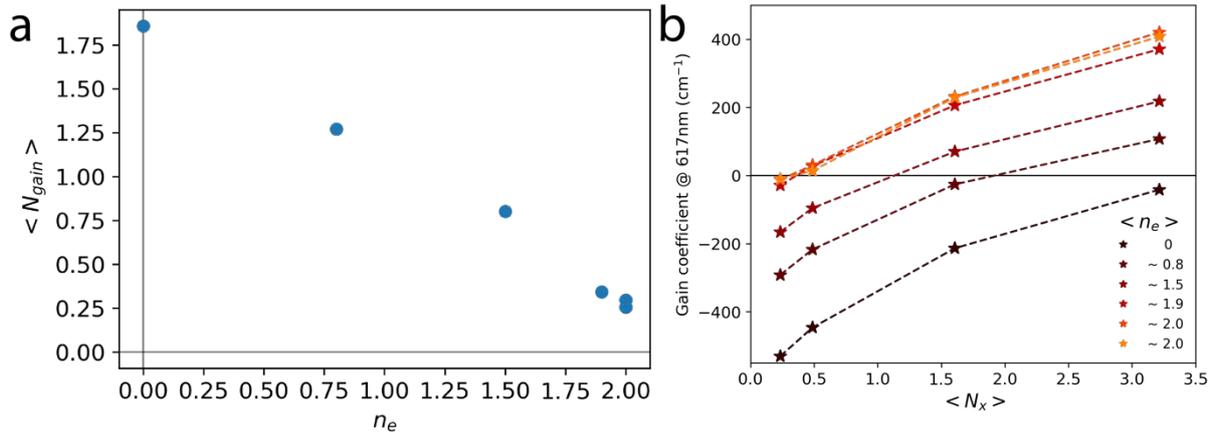

**Figure S41: Gain threshold as a function of doping density and gain coefficients.** (a) The threshold for optical gain is reduced from $<N_{gain}> = 1.8$ to $<N_{gain}> = 0.3$ upon doping the 1S(e) level with 2 electrons per QD. We hypothesize that the difference in gain threshold in these QDs and the QDs presented throughout the main text is due to the difference in PLQY, and hence related to trapping processes. Note that we effectively lower the optical gain threshold by 83%. (b) Gain coefficients at 617 nm (maximum gain), which reach a couple 100 cm$^{-1}$. These are slightly lower than the obtained gain coefficients for the QDs presented throughout the main text.

## Error calculations throughout the paper

In Figure 1(e); we calculate an average bleach -$\Delta A_{average}$ over a given wavelength range. The errorbars are ± two times the error in the mean ($\pm 2 \cdot \sigma_{mean}$) of the bleach in this wavelength range (i.e. the 95% confidence interval), calculated as $\sigma_{mean} = \sigma/\sqrt{n}$, where $\sigma$ is the standard deviation over the $\Delta A$ datapoints, and $n$ the number of datapoints over which the standard deviation (and -$\Delta A_{average}$) are calculated.

To obtain the gain threshold, we fit the |$\Delta A_{average}$| (or |$\Delta A$| for each wavelength in Figure 5(d,f)) versus <$N_X$> with an increasing exponential function; $|\Delta A_{average}| = A * (1 - e^{-B\langle N_X \rangle})$, where $A$ and $B$ are fit parameters. After fitting the curve, in order to obtain the gain threshold, we solve for which <$N_X$> the average bleach is equal to the average absorption value over the same wavelength range $A_0$;

$$f = \langle N_{X,gain} \rangle = -\frac{1}{B_{fit}} \ln\left(1 - \frac{A_0}{A_{fit}}\right)$$

The error in the gain threshold <$N_{X, gain}$> has to be propagated according to:

$$\sigma_{gain\ threshold} = \left(\frac{\partial f}{\partial A_{fit}}\right)^2 * \sigma_{A,1} + \left(\frac{\partial f}{\partial B_{fit}}\right)^2 * \sigma_{A,1} + 2\left(\frac{\partial f}{\partial A}\right)\left(\frac{\partial f}{\partial B}\right)\sigma_{AB,12}$$

Here, $\sigma_{A,1}$ and $\sigma_{B,1}$ the error in the the two fit parameters, i.e. the diagonal elements in the covariance matrix from the fit, and $\sigma_{AB,12}$ the off diagonal elements for both fit parameters (which describe the coupling between the errors in the fit parameters). Again, the displayed error bars on the gain threshold are the 95% confidence intervals, i.e. $\pm 2 \cdot \sigma_{gain\ threshold}$.

# SUPPLEMENTARY REFERENCES